\documentclass[a4paper,aps,11pt,nofootinbib]{revtex4}
\usepackage{bm,amsmath,amssymb}
\usepackage{graphicx}
\usepackage{paralist}
\usepackage{epstopdf}

\newcommand{\atanh}{\text{atanh}}

\begin{document}

\title{The Bethe approximation for solving the inverse Ising problem: a comparison with other inference methods}
\author{Federico Ricci-Tersenghi}
\affiliation{Dipartimento di Fisica, INFN--Sezione di Roma 1, and CNR--IPCF, UOS di Roma,
	Universit\`a La Sapienza, Piazzale Aldo Moro 5, I-00185 Roma, Italy}
\date{\today}

\begin{abstract}
The inverse Ising problem consists in inferring the coupling constants of an Ising model given the correlation matrix. The fastest methods for solving this problem are based on mean-field approximations, but which one performs better in the general case is still not completely clear.
In the first part of this work, I summarize the formulas for several mean-field approximations and I derive new analytical expressions for the Bethe approximation, which allow to solve the inverse Ising problem without running the Susceptibility Propagation algorithm (thus avoiding the lack of convergence). In the second part, I compare the accuracy of different mean field approximations on several models (diluted ferromagnets and spin glasses) defined on random graphs and regular lattices, showing which one is in general more effective. A simple improvement over these approximations is proposed. Also a fundamental limitation is found in using methods based on TAP and Bethe approximations in presence of an external field.
\end{abstract}

\maketitle

Mean-field approximations (MFA) are very important tools in statistical mechanics, since they provide an approximated description of a physical system in terms of few parameters (e.g.\ local magnetizations).
Among MFA the one based on the Bethe approximation (BA) is very effective. In recent years the BA --- originally derived for the ferromagnetic model on regular lattices \cite{Bethe} --- has been extended, under the name of Cavity Method, to models having arbitrary couplings and topologies \cite{MP}.
Although the BA is exact only for tree-like topologies, its application to models defined on random graphs has proved very successful: see e.g. the cases of low-density parity check codes, spin glasses and constraint satisfaction problems, all nicely reviewed in Ref.~\onlinecite{bookMarcAndrea}.

The inverse Ising problem, originally known as Boltzmann machine learning, consists in inferring coupling constants of an Ising model (both pairwise interactions and external fields) given the vector of magnetizations and the matrix of pairwise correlations.
In recent years the inverse Ising problem has received a lot of attention, specially in connection to inference in biological problems \cite{MartinPNAS, CoccoLeibler, PagnaniBMC, Roudi_FCN}.

The inverse Ising problems can be viewed as the dual problem with respect to the `direct' problem of estimating magnetizations and correlation given the Hamiltonian.
So, under any MFA, the inverse problem can be solved by inverting (if possible) the analytic expressions that give the magnetizations and the correlations as a function of interactions and fields.
Although MFA usually do not provide directly the correlations between distant variables, these correlations can be computed by using the linear response theorem \cite{KappenRodriguez}.

Within the BA, a very fast and efficient way to estimate correlations between any pair of variables is given by the Susceptibility Propagation (SuscProp) algorithm recently introduced in Ref.~\onlinecite{MM}. SuscProp is an iterative algorithm for solving a set of self-consistency equations: when it converges is very fast, but sometimes it may not converge.
Indeed the application of the BA to the inverse Ising problem has been limited up to now by the range of convergence of SuscProp \cite{EnzoValery,AurellOllion}.

In this work I present an analytical expression for the fixed point of SuscProp, thus avoiding any problem related to its lack of convergence. Actually such an expression already appeared in Ref.~\onlinecite{WellingTeh}, but was unknown to the statistical mechanics community (including the author): otherwise there would be no need for the SuscProp algorithm introduced in Ref.~\onlinecite{MM}.

In the first part of this work I derive new analytical expressions for solving the inverse Ising problem under the BA.
These analytical expressions, allow for a fair comparison among different MFA, in a wide range of temperatures, both for the problem of estimating 2-point correlations given the couplings (direct problem) and for the problem of estimating couplings given correlations and magnetizations (inverse problem).

In the present work I consider MFA obtained from the so-called Plefka expansion \cite{Plefka, GeorgesYedidia, Tanaka} and from a small correlations expansion \cite{SessakMonasson}. For the inverse Ising problem I compare methods that take in input only the correlation matrix. More complex, but usually slower, inference methods exist that require many samples of equilibrium configurations \cite{PLM, BentoMontanari, AurellEkeberg, CoccoMonasson} and not only the correlation matrix.

Improving over these MFA would be very welcome. It is well known that MFA ignore loops, so correcting these MFA by adding the loops contributions would be the right direction to follow. However at present all the methods which has been developed to consider explicitly the loops \cite{GBP, MontanariRizzo, MooijKappen_loop, LCBP2D, GBP-GF} do not provide analytical expressions for the correlations, which are simple enough to be inverted. For this reason, I have not considered these improved algorithms in the comparison of MFA for solving the inverse Ising problem.

Nonetheless, I am proposing a simple improvement of inference methods (both for the direct and the inverse problems) based on the idea that the loops may modify similarly self-correlations and correlations between close-by variables.

\section{The model and the mean-field approximations}
\label{sec:mfa}

In order to keep the presentation simple, I prefer to deal only with binary variables (Ising spins) $s_i=\pm1$ and Hamiltonian containing up to two-body interactions, i.e.\ external fields and pairwise couplings. Thus, the most general model I want to study is defined by the following joint probability distribution over $N$ Ising variables
\begin{equation}
P(s_1,\ldots,s_N) = \frac{1}{Z(\bm{J},\bm{h})} \exp\left[\sum_{i \neq j} J_{ij} s_i s_j + \sum_i h_i s_i\right]\;,
\label{measure}
\end{equation}
where the partition function $Z(\bm{J},\bm{h})$ is a normalizing constant, that depends on all the couplings $\bm{J} = \{J_{i,j}\}$ and the external fields $\bm{h} = \{h_i\}$. Please notice that the temperature parameter has been absorbed in the definition of external fields and couplings.
All the required information about the model is encoded in the free-energy
\begin{equation}
F(\bm{J},\bm{h}) = \ln Z(\bm{J},\bm{h})\;.
\end{equation}
In the rest of this Section I summarize the most common MFA to the free-energy: I am particularly interested in deriving the self-consistency equations for the magnetizations that are used in Section \ref{sec:corr} for obtaining 2-point correlations.

The simplest MFA, also known as naive MF (nMF), approximates the model in terms of local magnetizations $m_i=\langle s_i \rangle$, where the angular brackets represent the average w.r.t.\ the measure in Eq.(\ref{measure}).
The corresponding approximation to the free-energy is
\begin{equation}
F_\text{nMF} = \sum_i \left[ H\left(\frac{1+m_i}{2}\right) + H\left(\frac{1-m_i}{2}\right) \right] + \sum_i h_i m_i +\sum_{i \neq j} J_{ij} m_i m_j\;,
\end{equation}
where $H(x) \equiv -x \ln(x)$ and the $m_i$ must be fixed according to the self-consistency equations
\begin{equation}
\frac{\partial F_\text{nMF}}{\partial m_i} = \sum_j J_{ij} m_j + h_i - \atanh(m_i) = 0 \quad \Rightarrow
\quad m_i = \tanh\left[h_i + \sum_j J_{ij} m_j \right]\;.\label{miMF}
\end{equation}

A better MFA can be obtained by considering also the Onsager reaction term \cite{TAP}, leading to the following TAP approximated free-energy and self-consistency equations
\begin{eqnarray}
F_\text{TAP} &=& \sum_i \left[ H\left(\frac{1+m_i}{2}\right) + H\left(\frac{1-m_i}{2}\right) \right] + \nonumber \\
&& + \sum_i h_i m_i + \sum_{i \neq j} \left( J_{ij} m_i m_j + \frac12 J_{ij}^2 (1-m_i^2)(1-m_j^2) \right)\;, \label{FTAP}\\
m_i &=& \tanh\left[h_i + \sum_j J_{ij} \Big(m_j - J_{ij} (1-m_j^2) m_i\Big) \right]\;.\label{miTAP}
\end{eqnarray}
In the TAP approximation, when computing the marginal probability of spin $s_i$ (i.e.\ its magnetization $m_i$), the reaction term modifies the marginal probabilities of the neighboring spins, $m_j \to (m_j - J_{i,j} (1-m_j^2) m_i)$, in order to try to remove the effect of the spin $s_i$ under study. It has been recognized \cite{Plefka, GeorgesYedidia} that $F_\text{nMF}$ and $F_\text{TAP}$ are only the first two terms of the expansion of $F(\bm{J},\bm{h})$ in small couplings $\bm{J}$ at fixed magnetizations $\bm{m}=\{m_i\}$. This expansion contains \cite{GeorgesYedidia} both loop terms, like $J_{ij} J_{j\ell} J_{\ell i}$, and terms with higher powers of a single coupling, i.e.\ $J_{ij}^k$: the latter terms, that correspond to considering recursively the reaction to the reaction between spins $s_i$ and $s_j$, can be resummed and lead to the BA.

The BA gives a description of the model in terms of magnetizations $m_i$ and connected correlations $c_{ij}=\langle s_i s_j \rangle - m_i m_j$ between neighboring spins (i.e.\ spins connected by a non-zero coupling $J_{ij}$). The BA can be derived in two equivalent ways. The first way consists in finding values of $\bm{m}$ and $\bm{c}$ minimizing the following free-energy
\begin{align}
F_\text{BA} = \sum_{i \neq j} \left[ H\left(\frac{(1+m_i)(1+m_j)+c_{ij}}{4}\right) + H\left(\frac{(1-m_i)(1-m_j)+c_{ij}}{4}\right) + \right.\nonumber\\
+ \left. H\left(\frac{(1+m_i)(1-m_j)-c_{ij}}{4}\right) + H\left(\frac{(1-m_i)(1+m_j)-c_{ij}}{4}\right) \right] + \nonumber\\
+ \sum_i (1-d_i) \left[ H\left(\frac{1+m_i}{2}\right) + H\left(\frac{1-m_i}{2}\right) \right] + \sum_i h_i m_i + \sum_{i \neq j} J_{ij} (c_{ij} + m_i m_j)\;,\label{eq:FBA}
\end{align}
where $d_i$ is the degree of spin $s_i$, i.e.\ the number of its neighboring spins.
In Eq.(\ref{eq:FBA}) the last two terms correspond to the average value of the energy at given magnetizations and neighbouring correlations, while the first two terms correspond to the entropy of the Bethe approximation to the joint probability distribution of the $N$ spin variables,
\begin{equation}
P(s_1,\ldots,s_N) \stackrel{BA}{\simeq} \prod_{(ij)} \frac{p_{ij}(s_i,s_j)}{p_i(s_i)p_j(s_j)} \prod_i p_i(s_i)\;,
\end{equation}
where the first product runs over all pair of neighboring spins and the two-spins and single-spin marginal probabilities are given respectively by $p_{ij}(s_i,s_j) = [(1+m_i s_i)(1+m_j s_j) + c_{ij} s_i s_j]/4$ and $p_i(s_i) = (1+m_i s_i)/2$.
The conditions $\partial F_\text{BA} / \partial c_{ij} = 0$ can be solved analytically and lead to
\begin{eqnarray}
J_{ij} &=& \frac14 \ln\left(\frac{\Big((1+m_i)(1+m_j)+c_{ij}\Big)\Big((1-m_i)(1-m_j)+c_{ij}\Big)}{\Big((1+m_i)(1-m_j)-c_{ij}\Big)\Big((1-m_i)(1+m_j)-c_{ij}\Big)}\right)\;,\label{eq:IP}\\
c_{ij}(m_i,m_j,t_{ij}) &=& \frac{1}{2t_{ij}} \left(1+t_{ij}^2-\sqrt{(1-t_{ij}^2)^2-4t_{ij}(m_i-t_{ij}m_j)(m_j-t_{ij}m_i)}\right) - m_i m_j\;.\label{cijPrimo}
\end{eqnarray}
where $t_{ij} = \tanh(J_{ij})$.
Please note that Eq.(\ref{eq:IP}) is identical to Eq.(26) in Ref.~\onlinecite{SessakMonasson} and this is a further confirmation that resumming all 2-spin terms in the Plefka expansion leads to the BA.
Moreover Eq.(\ref{eq:IP}) has been used in the literature \cite{Roudi_FCN,Roudi_PRE} as the independent-pair (IP) approximation for inferring couplings from magnetizations and correlations: such an approximation infers the coupling $J_{ij}$ by assuming spins $s_i$ and $s_j$ form an isolated pair with magnetizations $m_i$ and $m_j$ and correlation $c_{ij}$. Unfortunately under this IP approximation computing the external fields in not immediate and moreover even the estimates of the couplings are rather poor (see Section \ref{sec:numeric}).

By making the substitution $c_{ij} \to c_{ij}(m_i,m_j,t_{ij})$ in $F_\text{BA}$ one can obtain the Bethe free-energy only in terms of magnetizations, from which the self-consistency equations for the magnetizations can be derived. However this derivation requires a rather complicated algebra and I prefer to obtain the same equations in a much simpler alternative way.

In the so-called Cavity Method \cite{MP} local magnetizations $m_i$ and neighbouring correlations $c_{ij}$ are expressed in terms of some auxiliary variables, the cavity magnetizations $m_i^{(j)}$ (i.e.\ the mean value of $s_i$ in the absence of a neighboring spin $s_j$):
\begin{eqnarray}
m_i &=& \frac{m_i^{(j)} + t_{ij}\, m_j^{(i)}}{1 + m_i^{(j)}\, t_{ij}\, m_j^{(i)}}\;,\label{mi}\\
m_j &=& \frac{t_{ij}\, m_i^{(j)} + m_j^{(i)}}{1 + m_i^{(j)}\, t_{ij}\, m_j^{(i)}}\;,\label{mj}\\
c_{ij} &=& \frac{t_{ij} + m_i^{(j)}\, m_j^{(i)}}{1 + m_i^{(j)}\, t_{ij}\, m_j^{(i)}} - m_i m_j\;.\label{cij}
\end{eqnarray}
Cavity magnetizations must satisfy the self-consistency equations
\begin{equation}
m_i^{(j)} = \tanh\left[h_i + \sum_{k (\neq j)} \atanh(t_{ik}\, m_k^{(i)})\right] \;.\label{SPEmij}
\end{equation}
These equations are often solved by an iterative algorithm known as Belief Propagation (BP) \cite{Pearl}: in case of convergence, the fixed point of BP gives directly the Bethe free-energy that admits an expression in terms of cavity magnetizations only \cite{MP}.

In order to obtain a closed set of self-consistency equations in the magnetizations $\bm{m}$, let me solve eqs.(\ref{mi}-\ref{mj}) for the cavity magnetizations and find
\begin{equation}
m_i^{(j)} = f(m_i, m_j, t_{ij}) \qquad m_j^{(i)} = f(m_j, m_i, t_{ij})\;,\label{mijmji}
\end{equation}
where
\begin{equation}
f(m_1, m_2, t) = \frac{1-t^2-\sqrt{(1-t^2)^2 - 4 t (m_1 - m_2 t)(m_2 - m_1 t)}}{2 t (m_2 - m_1 t)}\;.
\end{equation}
The sign in front of the square root has been chosen such that $f(0,0,t)=0$ as it should.
A consistency check can be made by substituting expressions (\ref{mijmji}) in Eq.(\ref{cij}) to obtain again the result in Eq.(\ref{cijPrimo}).
Finally, combining Eq.(\ref{mi}) and Eq.(\ref{SPEmij}), it is possible to obtain the self consistency equation for the magnetizations under the BA:
\begin{equation}
m_i = \tanh\left[ h_i + \sum_j \atanh\Big(t_{ij} f(m_j, m_i, t_{ij})\Big) \right]\;.\label{miBA}
\end{equation}
It is fair to comment that the use of this formula for finding Bethe magnetizations is not a good idea: indeed an iterative solution of Eq.(\ref{miBA}) is typically more unstable than BP solving Eq.(\ref{SPEmij}). My interest in this formula is that it involves only physical magnetizations (not cavity ones) and can be used to obtain correlations (see Section \ref{sec:corr}) and to solve in a fast way the inverse Ising problem (see Section \ref{sec:numeric}).

A series expansion of the exponent in Eq.(\ref{miBA}) for small couplings gives
\begin{equation}
h_i + \sum_j \atanh\Big(t_{ij} f(m_j,m_i,t_{ij})\Big) \simeq h_i + \sum_j \Big(J_{ij} m_j - J_{ij}^2 (1-m_j^2) m_i + \ldots\Big)\;,
\end{equation}
and one recognizes that the first two terms of the expansion are the naive MF approximation and the Onsager reaction term.
This expansion should make clearer that the BA is a way of considering recursively all the reactions between a pair of neighboring variables.

\section{Computing correlations by linear response}
\label{sec:corr}

A preliminary step to solve the inverse Ising problem by any MFA is to derive an analytical expression for the pairwise correlations as a function of the coupling constants.
Actually, the MFA discussed in Section \ref{sec:mfa} do not provide information about the correlation between distant variables: indeed, naive MF and TAP approximations give $c_{ij}=0$ for any pair of variables, and the BA only provides an expression for correlation between neighboring spins, see Eq.(\ref{cijPrimo}), which is trivially $c_{ij}=t_{ij}$ in case of null magnetizations.

Nonetheless, a closed set of equations for the connected correlations\footnote{Please do not confuse the correlation $C_{ij}$ with the parameter $c_{ij}$ appearing in the BA: the two coincide only when the BA is exact.}, $C_{ij} \equiv \langle s_i s_j \rangle - \langle s_i \rangle \langle s_j \rangle$ for any pair $i,j$, can be derived from the magnetizations self-consistency equations, Eqs.(\ref{miMF}), (\ref{miTAP}), (\ref{miBA}), through the linear response \cite{KappenRodriguez,WellingTeh}
\begin{equation}
C_{ij} = \frac{\partial m_i}{\partial h_j}\;, \qquad (C^{-1})_{ij} = \frac{\partial h_i}{\partial m_j}\;.
\end{equation}
The inverse correlation matrices $C^{-1}$ for the three MFA discussed above are given by the following expressions:
\begin{eqnarray}
\text{naive MF} \quad (C_\text{nMF}^{-1})_{ij} &=& \frac{\delta_{ij}}{1-m_i^2} - J_{ij}\;,\label{CinvMF}\\
\text{TAP} \quad (C_\text{TAP}^{-1})_{ij} &=& \left[\frac{1}{1-m_i^2}+\sum_k J_{ik}^2 (1-m_k^2)\right]\delta_{ij} - \left(J_{ij} + 2 J_{ij}^2 m_i m_j \right)\;,\label{CinvTAP}\\
\text{Bethe} \quad (C_\text{BA}^{-1})_{ij} &=& \left[\frac{1}{1-m_i^2}-\sum_k \frac{t_{ik}f_2(m_k,m_i,t_{ik})}{1-t_{ik}^2 f(m_k,m_i,t_{ik})^2}\right]\delta_{ij} - \frac{t_{ij}f_1(m_j,m_i,t_{ij})}{1-t_{ij}^2 f(m_j,m_i,t_{ij})^2}\;,\qquad\label{CinvBA}
\end{eqnarray}
where $f_1(m_1,m_2,t) \equiv \partial f(m_1,m_2,t) / \partial m_1$ and $f_2(m_1,m_2,t) \equiv \partial f(m_1,m_2,t) / \partial m_2$.
From these expressions one can obtain directly any correlation by simply computing the inverse of a matrix.

Please note that Eq.(\ref{CinvBA}) gives exactly the same solution found by the SuscProp iterative algorithm \cite{MM}, which is presently considered one among the best inference algorithms. The main advantage of Eq.(\ref{CinvBA}) is that it always provides the correlation matrix, even in those cases where SuscProp does not converge to the fixed point. Moreover inverting a matrix takes roughly the same time of a single iteration of SuscProp, and so using Eq.(\ref{CinvBA}) is much faster than running SuscProp, even when the latter converges.

Nevertheless, it is fair to notice that the use of Eq.(\ref{CinvBA}) does not solve all the problems related to the lack of convergence of SuscProp. Indeed, during the many tests I have run, I noticed that often the lack of convergence of SuscProp does correspond to the BA fixed point becoming unphysical: in these cases, by inverting the correlation matrix provided by Eq.(\ref{CinvBA}), one gets an unphysical correlation matrix (e.g. a correlation matrix with negative diagonal elements!). In this sense the lack of convergence of SuscProp gives a warning that the ``blind'' use of Eq.(\ref{CinvBA}) does not provide. So, a general suggestion when using the above formulas, providing an analytical expression for the correlation matrices under a MFA, is to check explicitly the physical consistency of the outcome.

One may comment that Eq.(\ref{CinvBA}) contains the magnetizations and the iterative computation of these (i.e.\ the BP algorithm) suffers the same convergence problems of SuscProp: this is easy to prove, given that the homogeneous SuscProp equations are nothing but the iterative equations for evolving under BP a small perturbation in the magnetization, and so BP is unstable if SuscProp does not converge. However there are provably convergent algorithms for the computation of magnetizations under the BA \cite{CCCP,HAK}: the use of these algorithms in conjunction with Eq.(\ref{CinvBA}) allows a direct computation of correlations under the BA. Moreover there are situations where magnetizations are known a priori and Eq.(\ref{CinvBA}) can be applied directly: e.g.\ when symmetries in the probability measure force magnetizations to be zero, or in the inverse Ising problem, where magnetizations are given as an input to the problem. In the rest of the paper I deal mainly with these two cases.

\subsection{Estimating correlations in case of null magnetizations}

A preliminary ranking of MFA can be done on the basis of how good are their estimates of correlations,  given the couplings. Indeed I expect that the better is this estimate, the better will be the solution to the inverse problem.

For simplicity I concentrate on models with no external fields and the couplings are multiplied by a parameter $\beta$ (the inverse temperature) such that the difficulty of the inference problem increases with $\beta$.

In case of null magnetizations, the expressions for the inverse correlation matrices simplify a lot
\begin{eqnarray}
\text{naive MF} \qquad (C_\text{nMF}^{-1})_{ij} &=& \delta_{ij} - J_{ij}\;,\label{first}\\
\text{TAP} \qquad (C_\text{TAP}^{-1})_{ij} &=& \left[1+\sum_k J_{ik}^2 \right]\delta_{ij} - J_{ij}\;,\\
\text{Bethe} \qquad (C_\text{BA}^{-1})_{ij} &=& \left[1+\sum_k \frac{t_{ik}^2}{1-t_{ik}^2}\right]\delta_{ij} - \frac{t_{ij}}{1-t_{ij}^2}\;,\label{CinvBAm0}
\end{eqnarray}
since $f_1(0,0,t)=1/(1-t^2)$ and $f_2(0,0,t)=-t/(1-t^2)$.

Given that for $m_i=0$ the expressions for the correlation matrices are much simpler, I report also those that can be obtained from the Plefka expansion at the third and fourth order
\begin{eqnarray}
3^\text{rd} \text{ order} \qquad (C_3^{-1})_{ij} &=& \left[1 + \sum_k J_{ik}^2 + 2 \sum_{k,\ell} J_{ik} J_{k\ell} J_{\ell i} \right]\delta_{ij} - \left( J_{ij} + \frac23 J_{ij}^3 \right)\;,\\
4^\text{th} \text{ order} \qquad (C_4^{-1})_{ij} &=& \left[1 + \sum_k J_{ik}^2 + 2 \sum_{k,\ell} J_{ik} J_{k\ell} J_{\ell i} + \frac13 \sum_k J_{ik}^4 + 2 \sum_{k,\ell,m} J_{ik} J_{k\ell} J_{\ell m} J_{mi} \right]\delta_{ij} + \nonumber\\
& & - \left( J_{ij} + \frac23 J_{ij}^3 + 2 J_{ij}^2 \sum_k J_{ik} J_{kj} \right)\;.\label{last}
\end{eqnarray}
The purpose is to understand if and how much does the estimate of the correlation matrix improve by adding terms in the Plefka expansion.

I have tested the accuracy of formulas in Eqs.(\ref{first}-\ref{last}) for ferromagnetic ($J_{ij}=1$) and spin glass ($J_{ij}=\pm1$) models defined on fully connected (FC) topologies, on a 2D square lattice and on a 3D cubic lattice. In diluted versions of these models a fraction $(1-p)$ of couplings has been set to zero. In models defined on FC graphs the couplings have been normalized such as to have a critical inverse temperature $\beta_c=1$ in the thermodynamic limit.

\begin{figure}
\includegraphics[width=0.9\columnwidth]{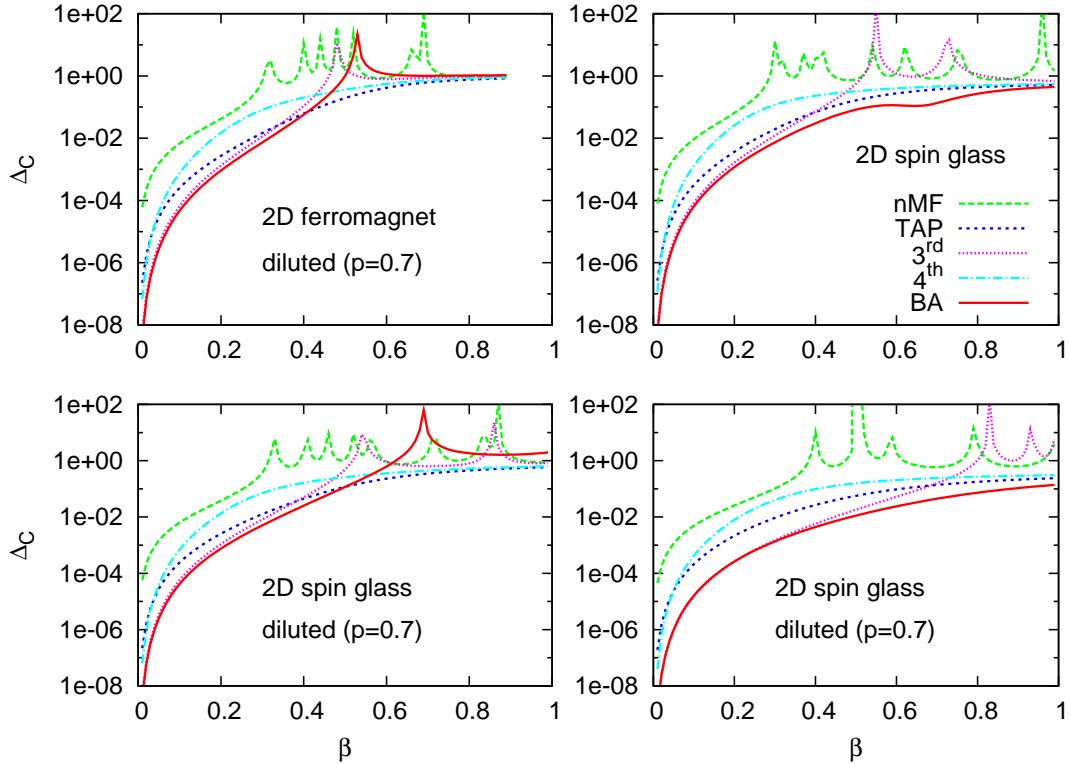}
\caption{Error made by 5 mean-field approximations in estimating the correlation matrix, given the couplings. Shown are typical samples of size $N=5^2$ (the qualitative behavior does not change for larger sizes).}
\label{plot_2D_all}
\end{figure}

\begin{figure}
\includegraphics[width=0.9\columnwidth]{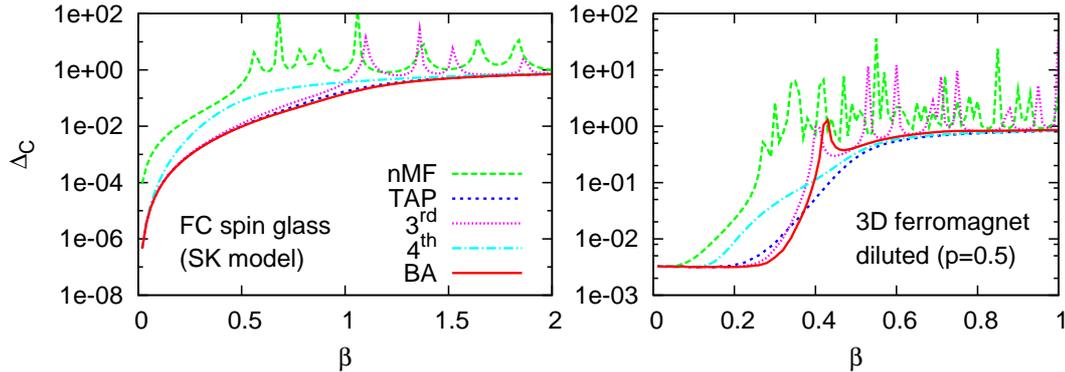}
\caption{Same as in Figure \ref{plot_2D_all} for typical samples of the fully-connected spin glass (SK model) of size $N=20$ and of the 3D diluted ferromagnet of size $N=3^5$. }
\label{plot_other}
\end{figure}

The discrepancy between true correlations $C$ and those inferred $C'$ is defined as
\begin{equation}
\Delta_C \equiv \sqrt{\frac{1}{N^2} \sum_{i,j} (C_{ij} - C'_{ij})^2}\;.
\end{equation}
In Figure \ref{plot_2D_all} and \ref{plot_other} I report the typical behavior of the error $\Delta_C$ between exact and estimated correlation matrices for 5 different MFA. 
Figure \ref{plot_2D_all} shows results for models defined on a 2D square lattice, while Figure \ref{plot_other} refers to FC and 3D topologies.
In order to compare the MFA estimates with the exact correlation matrices I am studying here small systems, but the qualitative behavior does not change for larger sizes.

Although the quantitative behavior of $\Delta_C$ depends on the specific sample, some general statements can be made:
\begin{compactitem}
\item naive MF is typically the worst MFA and shows many spurious singularities (roughly one for each peak in $\Delta_C$);
\item TAP and $4^\text{th}$ order approximations typically show no (or very rare) singularities;
\item the best estimate is typically provided by BA and TAP, with BA being the best unless it has a singularity (in this case TAP becomes the best at lower temperatures, higher $\beta$).
\end{compactitem}
These results suggest that increasing the number of terms in the Plefka expansion does not always improve the estimate of the correlation matrix (as one could have naively expected).
On the basis of these preliminary results I believe it is relevant to consider only TAP and BA for the inverse Ising problem, together with other inference methods (see Section \ref{sec:inverse}).

In the left panel of Figure \ref{plot_other} the results obtained by TAP and BA are almost perfectly superimposed (indeed the former in not well visible). This is expected since the TAP approximation is exact for the SK model at high temperatures ($\beta < \beta_c = 1$) in the large $N$ limit: so the BA can not improve it, but in $1/N$ corrections. Indeed a careful analysis shows a tiny improvement of BA over TAP around the critical temperature, where $1/N$ corrections are stronger.

Please note that in the right panel of Figure \ref{plot_other} the high temperature (small $\beta$) behavior of $\Delta_C$ is very different than in previous plots: indeed for $\beta \to 0$, $\Delta_C$ goes to a constant, instead of decreasing with a power law in $\beta$ (as in Figure \ref{plot_2D_all} and in the left panel of Figure \ref{plot_other}). This is due to the fact that the comparison has not been made with the exact correlation matrix, but with correlations measured from a Monte Carlo (MC) simulation. Actually, in this case, I have used the Wolff algorithm and the correlation matrix has been computed from $10^5$ independent measures.
The difference between the error due to the MFA and the error due to MC noisy data can be better appreciated in Figure \ref{plot_2D_MCvsExact}: in the high temperature region the error does not decrease below a limiting value given roughly by the inverse of the square root of the number of measures.

\section{Improving inference algorithms}
\label{sec:trick}

Expressions in Eqs.(\ref{first}-\ref{last}) are intrinsically approximated, and turn out to be correct only in some particular cases. Naive MF and TAP approximations (as well as $3^\text{rd}$ and $4^\text{th}$ orders approximations), being the first orders in a small couplings expansion, are exact only in the limit of very weak couplings (either high temperature or fully-connected models in the large $N$ limit). The BA, on the contrary, is exact also for coupling intensities $O(1)$, but only if the interacting network is a tree; on random graph models (which are locally tree-like) the BA turns out to be correct as long as the model has only one state (modulo the known symmetries). On any other model those expressions are approximated and it is worth trying to improve it.

Let me first notice that any of the above MFA returns in general a value for the self-correlation differing from the exact one, i.e.\ $C_{ii} \neq 1$ (for simplicity I consider the case of null magnetizations, but the argument is general). This fact can be easily explained, noticing that all the above MFA assume that correlations along loops are vanishingly small (at least in the large $N$ limit). On the contrary, on any loopy graph, like e.g.\ regular lattices, correlations along loops are important and may alter significantly the mean-field estimates. A general solution to this problem is still not available, although many work is in progress to include loop contributions to MFA \cite{GBP,MooijKappen_loop,LCBP2D,GBP-GF,MontanariRizzo}.

What I am proposing here is a simple heuristic improvement. Once the correlation matrix $C_{ij}$ is computed by one of the approximations described in Section \ref{sec:corr}, a properly \emph{normalized} correlation matrix can be defined
\begin{equation}
\widehat{C}_{ij} \equiv \frac{C_{ij}}{\sqrt{C_{ii}C_{jj}}}\;.
\label{normC}
\end{equation}
By definition $\widehat{C}_{ii}=1$ and also off-diagonal element may approximate better the true correlations. The reason for this is that the loops neglected in MFA actually modify in a similar way both self-correlations $C_{ii}$ and off-diagonal correlations $C_{ij}$, and the heuristic normalization in Eq.(\ref{normC}) is assuming that the modifying factor only depends on the loop structure around sites $i$ and $j$ (which is certainly wrong for distant sites, but may be a reasonable approximation for closed-by sites).

\begin{figure}
\includegraphics[width=0.7\columnwidth]{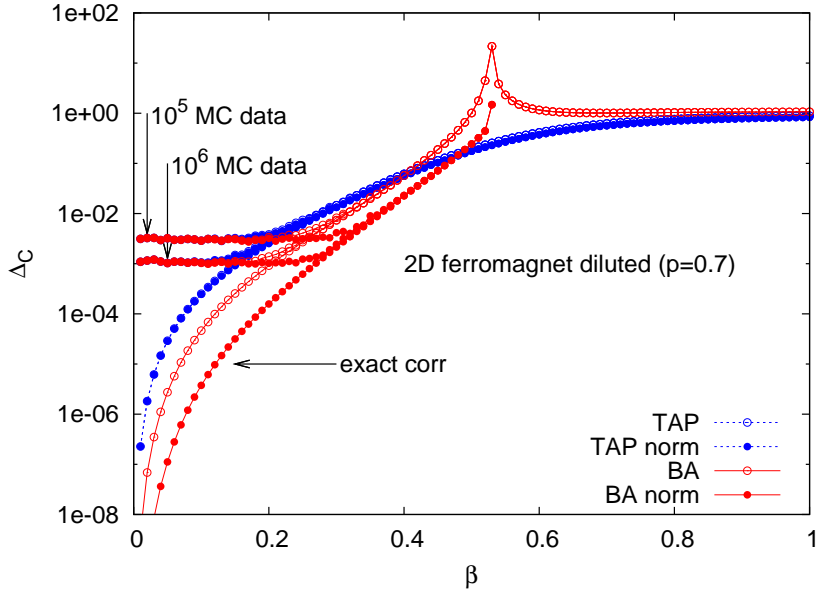}
\caption{Same as in Figure \ref{plot_2D_all} for a typical sample of the 2D diluted ferromagnet of size $N=5^2$. The error $\Delta_C$ has been computed with respect to the exact correlation matrix and with respect to the one measured in MC simulations. Full points show the error obtained with the normalization trick.}
\label{plot_2D_MCvsExact}
\end{figure}

In Figure \ref{plot_2D_MCvsExact} full points show that the error $\Delta_C$ in the BA decreases by roughly one order of magnitude if normalized correlations are used. On the contrary, the TAP result is not very sensitive to this normalization: the reason is that the estimates of the self-correlations in TAP remain quite close to the right value, specially if compared to BA estimates that diverge at the singularity (mark by a peak in Figure \ref{plot_2D_MCvsExact}). On the right of such a peak the error obtained by the normalized BA is not reported because Eq.(\ref{normC}) can not be used, since several BA estimates for self-correlations are negative. This is the problem of the BA fixed point becoming (strongly) unphysical, already discussed in Section \ref{sec:corr}: indeed by running SuscProp on this sample one would observe convergence only for $\beta$ smaller than the peak location. I would like to stress again that checking the physical consistency of a solution based on a MFA is very important: for the sample shown in Figure \ref{plot_2D_MCvsExact}, even without knowing the exact correlations, one should switch from the BA to TAP, when the former reaches the singularity (that manifests e.g.\ in SuscProp not converging or in self-correlations diverging)\footnote{Actually for a ferromagnet one knows how to break the up-down symmetry and let BP converge even at low temperatures: once BP returns non-zero magnetizations $m_i$, the correlation matrix can be computed by mean of Eq.(\ref{CinvBA}). However in the general case, BP does not converge in presence of long range correlations, i.e.\ after the singularity, and one must resort to other MFA.}.

Moreover there are cases (e.g. homogeneous FC models) where the spurious singularity induced by the MFA in a system of finite size is such that $C_{ii}$ and $C_{ij}$ diverge with the same law at the spurious critical point, while the normalized correlation $\widehat{C}_{ij}$ stays finite (and much closer to the true one). For example for the FC ferromagnetic model the normalized correlation $\widehat{C}_\text{MFA}$ estimates the true correlation with an error roughly half than the one of $C_\text{MFA}$ for any of the 5 MFA considered here.

\section{Methods for the inverse Ising problem}
\label{sec:inverse}

I consider 4 different approximations for solving the inverse Ising problem.
The simplest one is the independent-pair (IP) approximation, already discussed in Section \ref{sec:mfa} and recalled here for convenience
\begin{equation}
J^\text{IP}_{ij} = \frac14 \ln\left(\frac{\Big((1+m_i)(1+m_j)+C_{ij}\Big)\Big((1-m_i)(1-m_j)+C_{ij}\Big)}{\Big((1+m_i)(1-m_j)-C_{ij}\Big)\Big((1-m_i)(1+m_j)-C_{ij}\Big)}\right)\;.
\label{IP}
\end{equation}
Among the MFA which can be derived from the Plefka expansion, I consider only TAP and BA, because are those performing better in the direct problem of estimating correlations (see Section \ref{sec:corr}). The corresponding expressions for the inferred couplings can be obtained by solving the equation
\begin{equation}
2 m_i m_j J_{ij}^2 + J_{ij} +(C^{-1})_{ij} = 0\quad\forall (i \neq j)
\label{eq:TAP}
\end{equation}
for TAP and the equation
\begin{equation}
(C^{-1})_{ij} = \frac{-t_{ij}f_1(m_j,m_i,t_{ij})}{1-t_{ij}^2 f(m_j,m_i,t_{ij})^2} = \frac{-t_{ij}}{\sqrt{(1-t_{ij}^2)^2-4t_{ij}(m_i-t_{ij}m_j)(m_j-t_{ij}m_i)}}\quad\forall (i \neq j)
\label{eq:BA}
\end{equation}
for the BA, thus leading to
\begin{eqnarray}
J^\text{TAP}_{ij} &=& \frac{\sqrt{1-8 m_i m_j (C^{-1})_{ij}} - 1}{4 m_i m_j}\;,\label{TAP}\\
J^\text{BA}_{ij} &=& -\atanh\Bigg[\frac{1}{2(C^{-1})_{ij}}\sqrt{1+4(1-m_i^2)(1-m_j^2)(C^{-1})^2_{ij}}-m_i m_j\,-\nonumber\\
&&\frac{1}{2(C^{-1})_{ij}}\sqrt{\left(\sqrt{1+4(1-m_i^2)(1-m_j^2)(C^{-1})^2_{ij}}-2 m_i m_j(C^{-1})_{ij}\right)^2-4(C^{-1})^2_{ij}}\;\Bigg]\;.\label{BA}
\end{eqnarray}
The fourth approximation I am considering has been obtained from a small correlation expansion by Sessak and Monasson \cite{SessakMonasson} and has been further simplified in Ref.~\onlinecite{Roudi_PRE} to the following expression
\begin{equation}
J^\text{SM}_{ij} = -(C^{-1})_{ij} + J^\text{IP}_{ij} - \frac{C_{ij}}{(1-m_i^2)(1-m_j^2)-(C_{ij})^2}\;.
\end{equation}

For each approximation, I measure the error in inferred couplings $J'_{ij}$ with respect to the true ones $J_{ij}$ by the following expression
\begin{equation}
\Delta_J = \sqrt{\frac{\sum_{i<j} (J'_{ij} - J_{ij})^2}{\sum_{i<j} J_{ij}^2}}\;.
\end{equation}

I study both diluted ferromagnetic model with a fraction $p$ of non-zero couplings ($J_{ij}=\beta$) and undiluted spin glass models ($J_{ij} = \pm\beta$ with probability $1/2$). I also consider several topologies: 2D square lattices, 3D cubic lattices, random regular graphs with fixed degree $c=4$ and fully connected (FC) graphs. In the latter case the couplings are normalized in order to have a phase transition at $\beta_c=1$ in the thermodynamic limit. I restrict the study to models of small sizes, with $N$ ranging between 20 and 100, because these are the sizes for problems of biological interest. Moreover, as discusses below, the number $M$ of independent measurements of the correlation matrix that make inferred coupling reasonably good grows linearly with $N$ and so for larger systems the number of measurements needed become too large.
The data shown in Section~\ref{sec:numeric} have been obtained with $M=10^6$ independent measures of the correlation matrix (unless differently stated) and going to much larger values seems to me rather unrealistic if compared with practical applications.

\subsection{Normalization trick for the inverse Ising problem}

The trick of normalizing the correlation matrix to improve inference (see Section \ref{sec:trick}) can be extended to the inverse Ising problem. In practice, it corresponds to solve \emph{all} the equations relating the inverse correlation matrix $C^{-1}$ to the couplings $J_{ij}$, including also those for the diagonal elements which are usually ignored.

Let me illustrate the new method for the simple case of the TAP approximation with null magnetizations.
In this case, solving the inverse Ising problem only on the off-diagonal elements is equivalent to solve the equations
\[
(C^{-1})_{ij} = - J_{ij} \equiv D_{ij} \quad \forall (i \neq j)\;,
\]
but the diagonal equations are in general unsatisfied
\[
(C^{-1})_{ii} \neq 1 + \sum_k J_{ik}^2 \equiv D_{ii}\;,
\]
where $D$ is the inverse correlation matrix estimated by TAP once coupling $J_{ij}$ are given.
Please notice that diagonal elements $D_{ii}$ are fully determined once the off-diagonal elements are known.

Following Eq.(\ref{normC}) I would like to normalize the estimated matrix $D$, and produce a normalized inverse correlation matrix $\widehat{D}$, matching better the true inverse correlation matrix $C^{-1}$. In other words I would like to solve the equations
\begin{equation}
(C^{-1})_{ij} = \widehat{D}_{ij} = D_{ij} \lambda_i \lambda_j \quad \forall i,j\;,
\end{equation}
where the $N$ variables $\lambda_i$ are exactly those necessary to solve the new $N$ diagonal equations.
Physically speaking, $\lambda_i^2$ should be the self-correlation $C_{ii}$ produced by the MFA, when using the right couplings $J_{ij}$.
When the $\lambda_i$ becomes very different from 1, then the MFA is working very badly; however I expect situations where some of the errors produced by the MFA can be compensated by this normalization trick.

Unfortunately the solution to the new equations, those involving both $J_{ij}$ and $\lambda_i$, does not have an analytical expression and need to be solved numerically. I have adopted an iterative solution, which is very fast (when it converges).
In practice, I start with all $\lambda_i=1$ and then, iteratively, first I solve the off-diagonal equations, thus getting an estimate for the couplings, and then I solve the diagonal equations to obtain a new estimate for the $\lambda$'s. I repeat this iterative procedure, updating $\lambda$'s with a damping factor, until the $\lambda$'s variations is below a threshold (typically $10^{-8}$).

From the many tests I have run, I noticed that this normalization trick is more relevant for  unfrustrated models (as the diluted ferromagnets studied below) or models containing regions very weakly frustrated (those usually leading to Griffith singularities).
Most probably in these weakly frustrated regions correlations get self-reinforced by the loops (ignored in the MFA) and thus the normalization trick may improve coupling estimates.
On the contrary for strongly disordered models, like spin glasses, the effect of the normalization trick depends a lot on the disorderd sample and it does not seem to give a clear improvement on average.

The use of the normalization trick for improving inference in the inverse Ising problem may resemble the diagonal-weight trick introduced in Ref.~\onlinecite{KappenRodriguez}, but it is actually very different. In the diagonal-weight trick, the self-couplings $J_{ii}$ are allow to take non-zero values in order to solve all the equations $(C^{-1})_{ij}=D_{ij}$, while in the normalization trick the self-couplings $J_{ii}$ remain null. It has been shown \cite{Tanaka} that the the first order approximation (nMF) with the diagonal-weight trick improves over the second order approximation (TAP) in estimating magnetizations. However the estimates for the couplings do not improve at all, because the off-diagonal equations are left unchanged by the diagonal-weight trick. On the contrary, the normalization trick used here does change the estimates for the couplings. Moreover, being based on the very general requirement that self-correlations must take the right value, it can be applied to any approximation.

\section{Numerical results on the inverse Ising problem}
\label{sec:numeric}

Let me start by making some general statements that summarize the numerical results shown in this Section. Among the four approximations studied (IP, TAP, SM and BA) it seems in general true that:
\begin{itemize}
\item IP always provides the worst estimate, especially at high temperatures (low $\beta$);
\item BA always outperforms TAP;
\item between SM and BA, the former is better in the high temperature (low $\beta$) phase, while the latter is better at lower temperatures (higher $\beta$ values);
\item at low $\beta$, the error $\Delta_J$ in couplings inference is completely dominated by the uncertainty on the correlation matrix and it does not depend on the inference method: for this reason the range where SM is the best method becomes tiny, especially for noisy data;
\item at high $\beta$, the error $\Delta_J$ produced by TAP and SM diverge, the one by IP stays limited, but very high, and only BA may have a reasonable error;
\item for diluted ferromagnetic models the normalization trick works fine and thus BA with the normalization trick is the best method overall;
\item in presence of an external field, i.e.\ when magnetizations are different from zero, TAP and BA stop working at high enough $\beta$ values (i.e.\ these methods do not admit a solution); given that at high $\beta$ neither IP nor SM provide acceptable inferred couplings, I conclude that in such a situation the inverse Ising problem needs to be solved by other methods not explored in the present work.
\end{itemize}

\subsection{Ferromagnetic models}

\begin{figure}
\includegraphics[width=0.6\columnwidth]{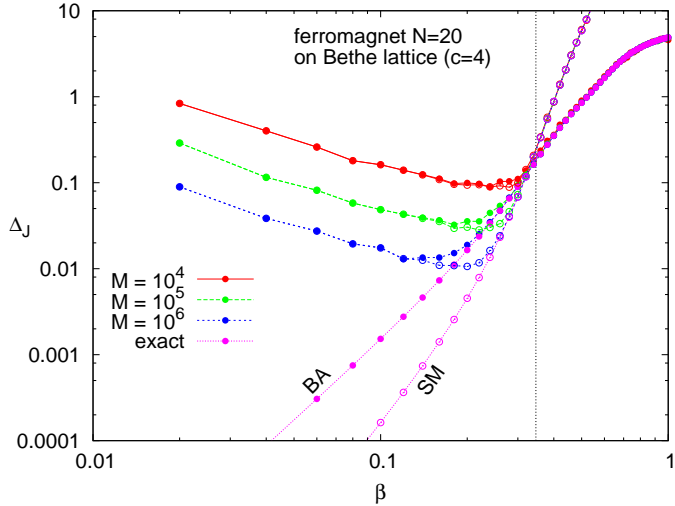}
\caption{Dependence of the error $\Delta_J$ in inferred couplings on the number $M$ of independent measures for the correlation matrix. The model is a ferromagnet of $N=20$ variables on a regular random graph of fixed degree $c=4$, whose critical temperature is marked by the vertical dotted line.}
\label{fig:numSam}
\end{figure}

Let me start discussing the high temperature (low $\beta$) regime. In Figure~\ref{fig:numSam} are reported the errors $\Delta_J$ in inferring the couplings of a ferromagnetic model of $N=20$ variables on a regular random graph of fixed degree $c=4$. For simplicity I am plotting only the results obtained with SM and BA. Data in the upper curves have been computed using a correlation matrix averaged over $M$ independent measurements, while data in lower curves have been computed from the exact correlation matrix. It is clear that the low $\beta$ regime is completely dominated by the uncertainty in the correlation matrix (as already noticed in Ref.~\cite{EnzoValery}), and the error in this regime is independent on the inference method used. For this reason the good performances of SM in this regime are actually washed out and, even for $M=10^6$ measurements, the improvement of SM over BA is very limited (see Figure~\ref{fig:numSam}). Moreover such an improvement tends to become smaller by increasing the system size because in the low $\beta$ regime the error goes like
\begin{equation}
\Delta_J \propto \frac{1}{\beta}\sqrt{\frac{N}{M}}\;.
\end{equation}

I think that comparing inference methods by using the exact correlation matrix is rather unrealistic, given that in any practical application the correlations are always known with some uncertainty. So in presenting below the numerical results I always consider the case with $M=10^6$ independent measurements for the magnetizations and the correlations.

\begin{figure}
\includegraphics[width=0.49\columnwidth]{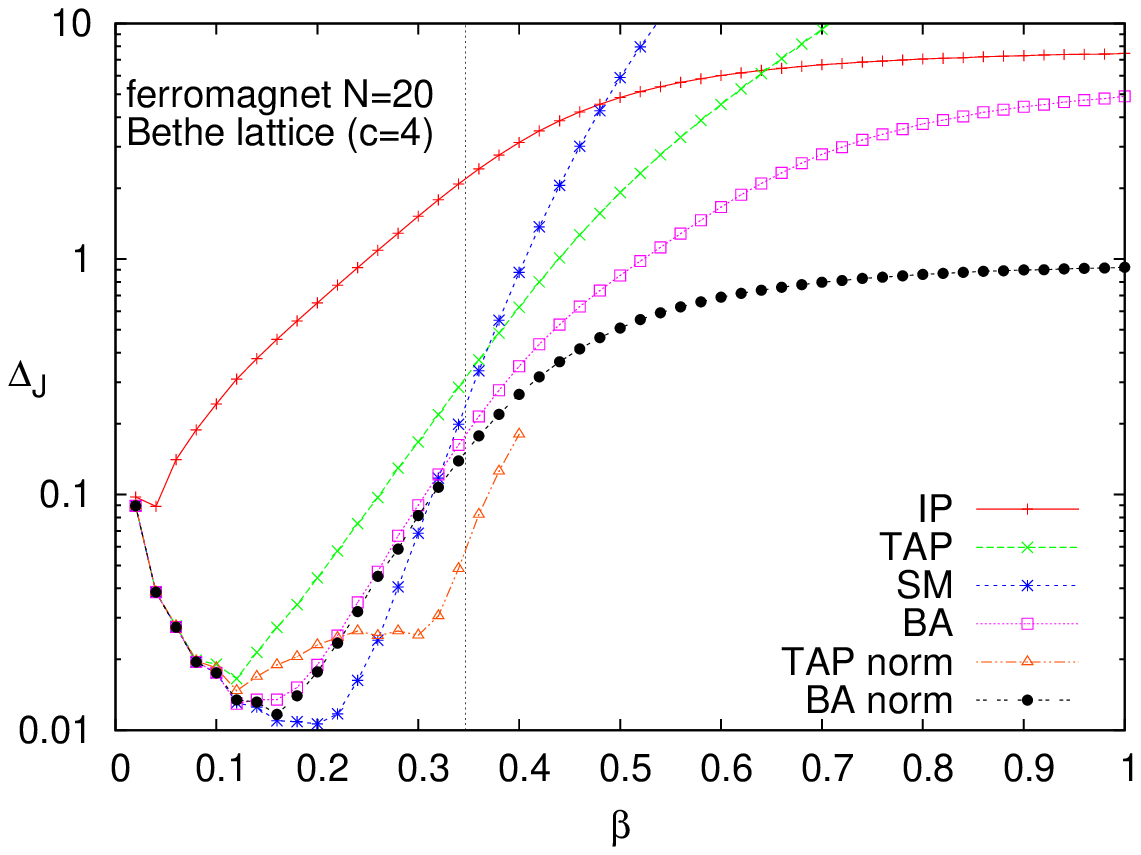}
\includegraphics[width=0.49\columnwidth]{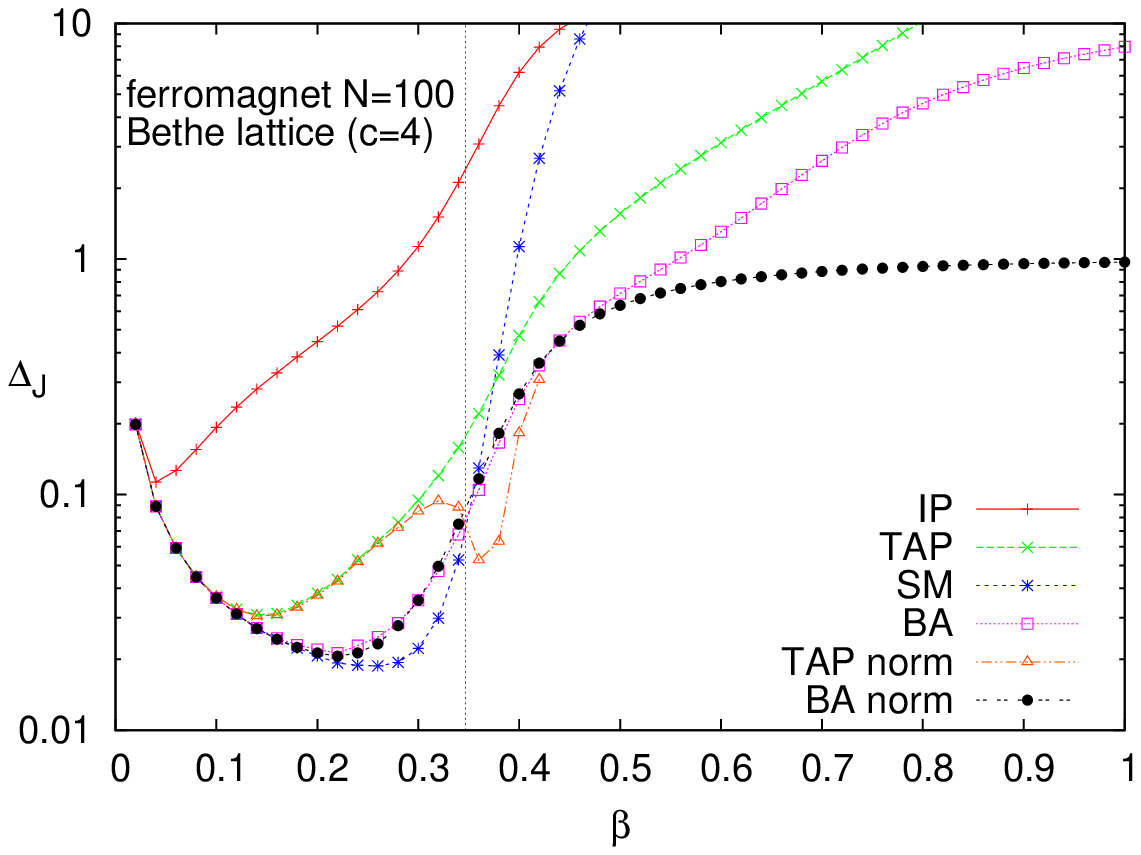}
\caption{Errors in the couplings inferred by several approximations. The model is a ferromagnet on a random regular graph of fixed degree 4. Shown are two typical samples of sizes $N=20$ (left) and $N=100$ (right). The vertical dotted lines mark the locus of the ferromagnetic phase transition in the thermodynamic limit.}
\label{ferroBethe}
\end{figure}

\begin{figure}
\includegraphics[width=0.4\columnwidth]{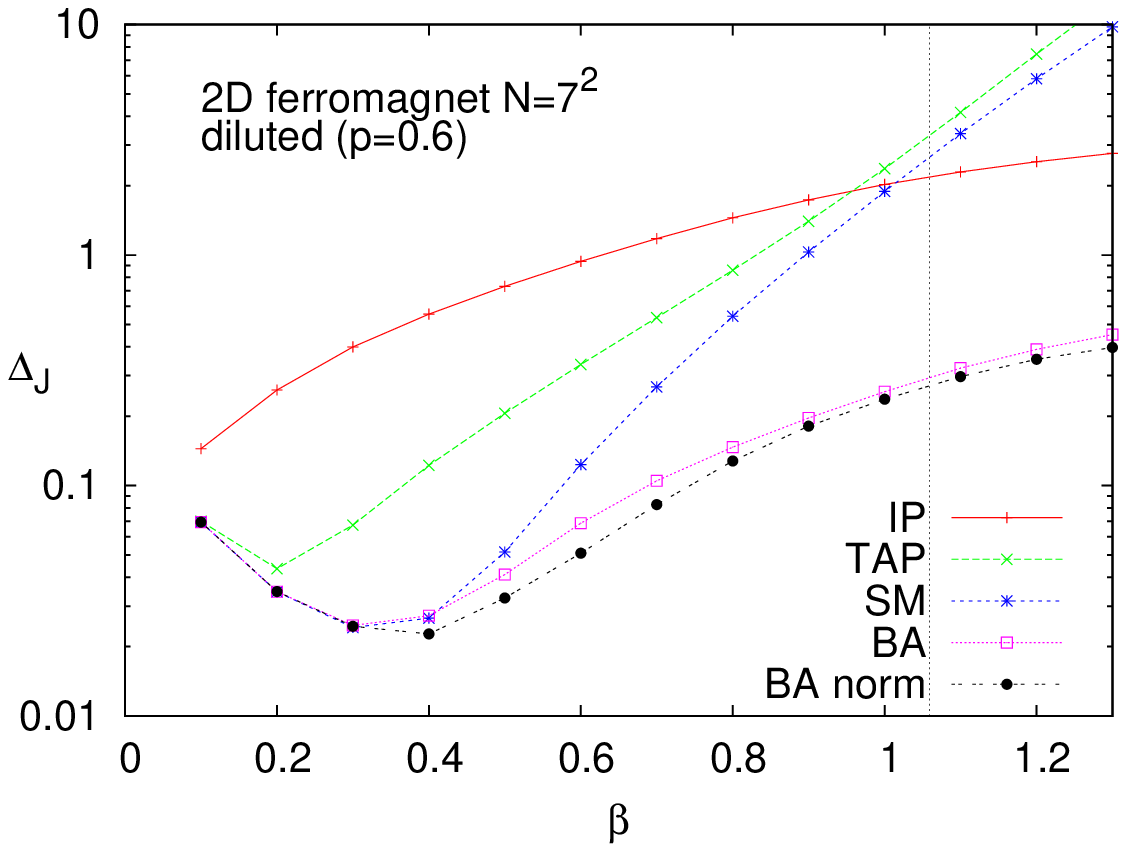}
\includegraphics[width=0.4\columnwidth]{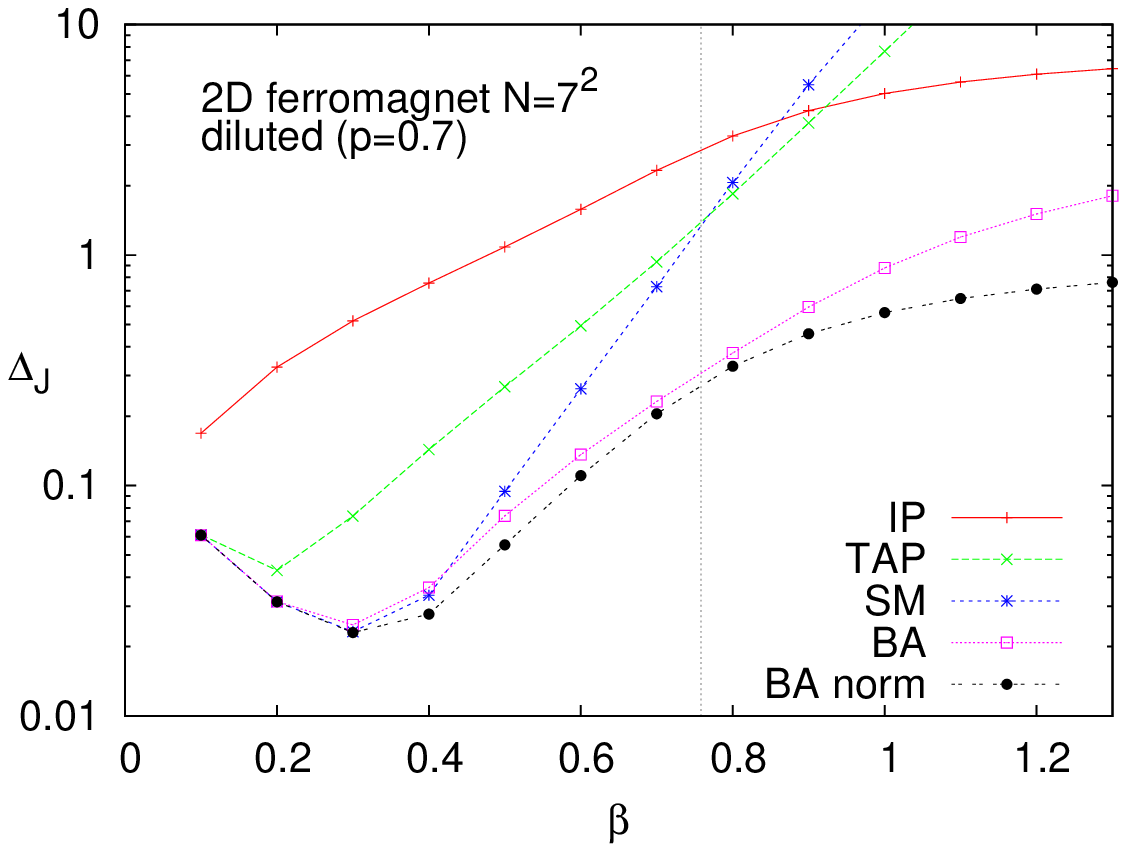}
\includegraphics[width=0.4\columnwidth]{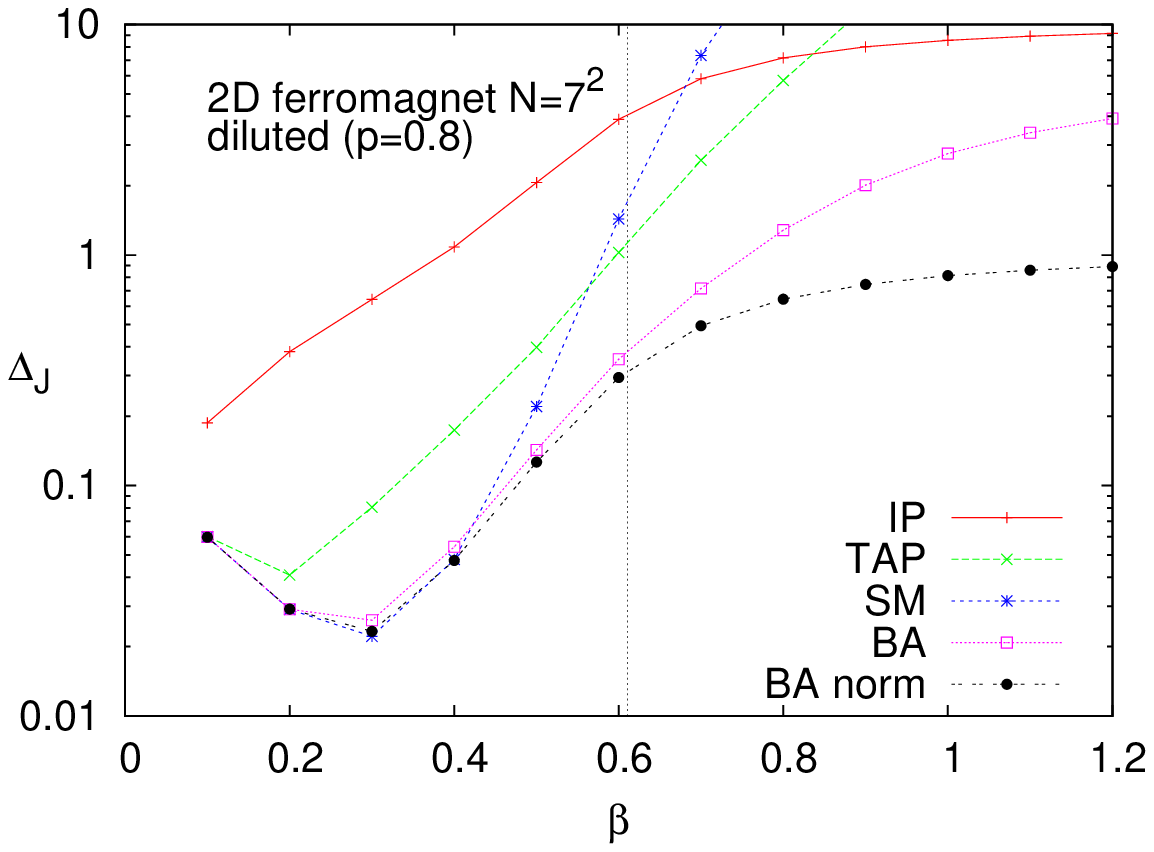}
\includegraphics[width=0.4\columnwidth]{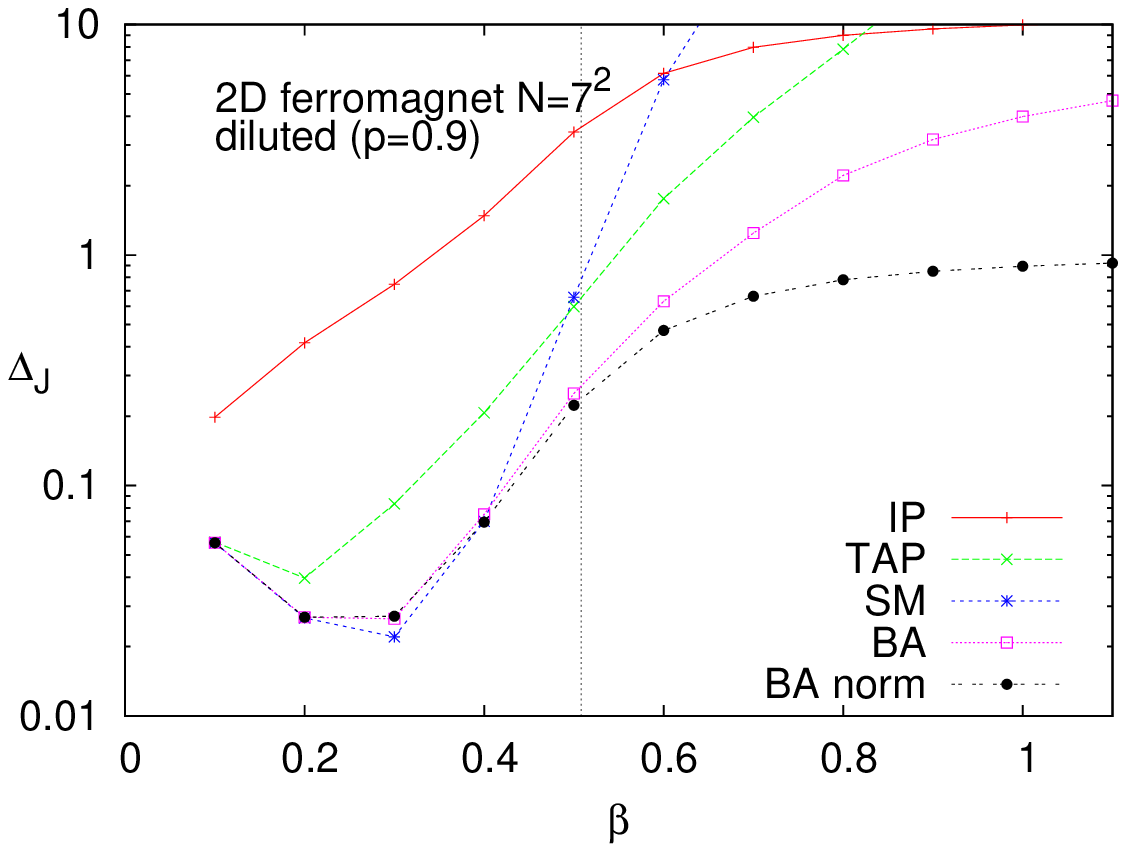}
\caption{Errors in the couplings inferred by several approximations. The model is a diluted ferromagnet on a 2D square lattice of size $N=7^2$. Shown are typical samples for 4 different dilutions: 0.6, 0.7, 0.8 and 0.9. The vertical dotted lines mark the loci of the ferromagnetic phase transitions in the thermodynamic limit.}
\label{ferro2D}
\end{figure}

In Figure \ref{ferroBethe} I am showing the error in the couplings inferred by several approximation for a ferromagnet on a random regular graph with fixed degree $c=4$. The two panels correspond to sizes $N=20$ (left) and $N=100$ (right) and show that the qualitative behavior is mostly size independent. Also the dependence on the specific sample (i.e., on the random graph) is rather weak. The data in Figure~\ref{ferroBethe} support many of the statements written above: (i) IP is a very bad approximation even in the low $\beta$ regime; (ii) BA always outperforms TAP; (iii) SM is better than BA only in the low $\beta$ regime, but here the error is dominated by the uncertainty in the correlation and increases with the system size, so the improvement of SM over BA is tiny; (iv) errors in TAP and SM diverge for large $\beta$, while those in IP and BA remains finite, although very large; (v) the normalization trick works nicely and gives actually the best result in a wide range of temperatures. The data for TAP with the normalization trick (label ``TAP norm'') are interrupted because at large $\beta$ the iterative procedure I am using for finding the parameters $\{\lambda_i\}$ stops converging.

The same qualitative conclusions can be reached by studying a diluted ferromagnet on a 2D square lattice for several different dilutions (see Figure~\ref{ferro2D}). In particular the relative quality of the approximations seems to be independent on the dilution, and the BA with the normalization trick outperforms the other inference methods. However I notice that, while the error of BA (with and without the normalization trick) at the critical temperature is roughly independent on the dilution, the errors made by TAP and SM tend to increase when the dilution is stronger and the system becomes more heterogenous.

\begin{figure}
\includegraphics[width=0.49\columnwidth]{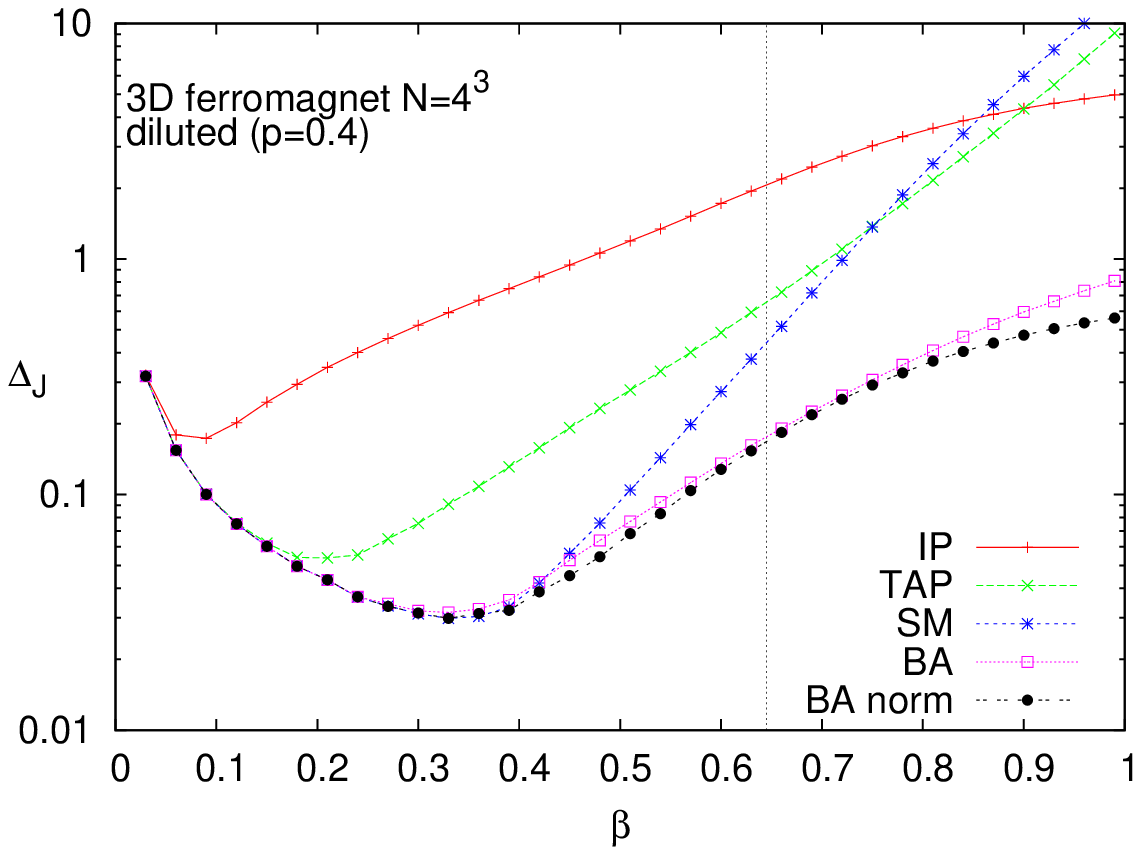}
\includegraphics[width=0.49\columnwidth]{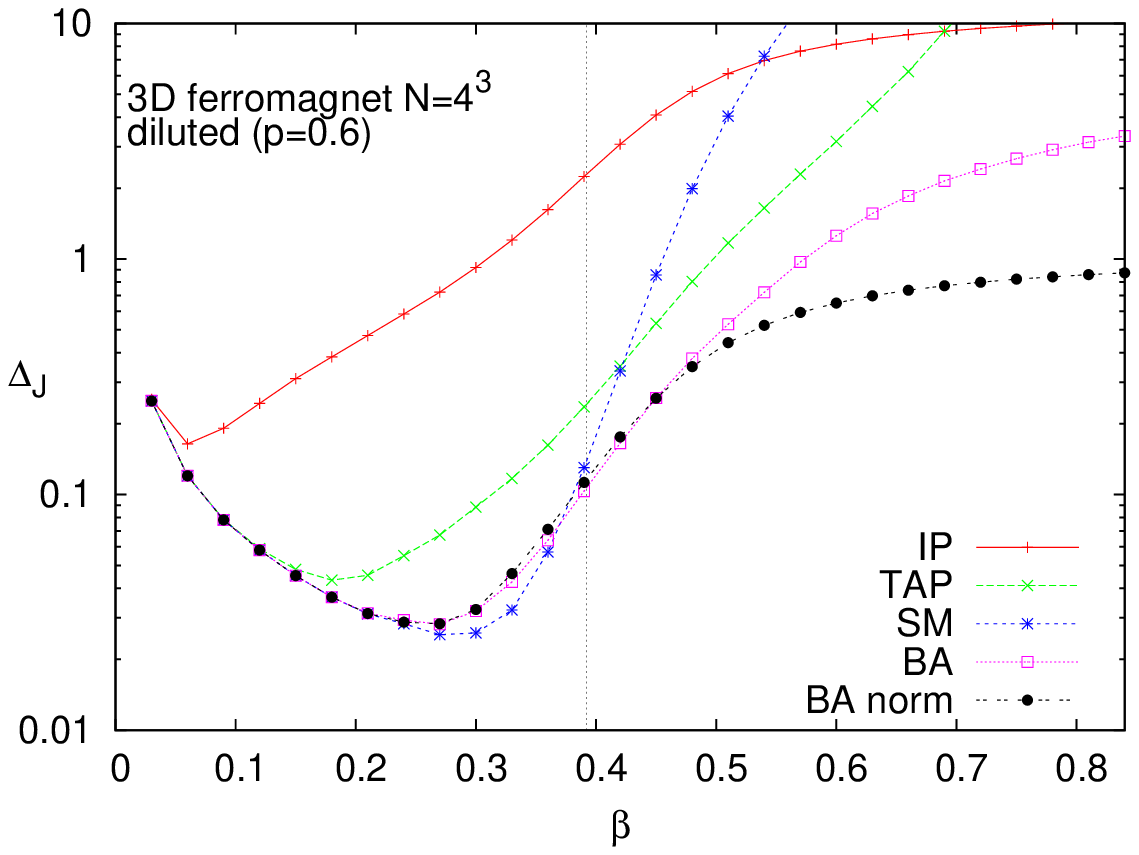}
\caption{Errors in the couplings inferred by several approximations. The model is a diluted ferromagnet on a 3D cubic lattice of size $N=4^3$. Shown are two typical samples with different dilutions. The vertical dotted lines mark the loci of the ferromagnetic phase transitions in the thermodynamic limit.}
\label{ferro3D}
\end{figure}

All the conclusions reached for the 2D case, perfectly apply also to the case of a diluted ferromagnetic model on a 3D cubic lattice (see Figure~\ref{ferro3D}). So it is very reasonable to conclude that the statements made at the beginning of this Section apply to any diluted ferromagnet independently on the specific topology.

\subsection{Spin glass models}

\begin{figure}
\includegraphics[width=0.49\columnwidth]{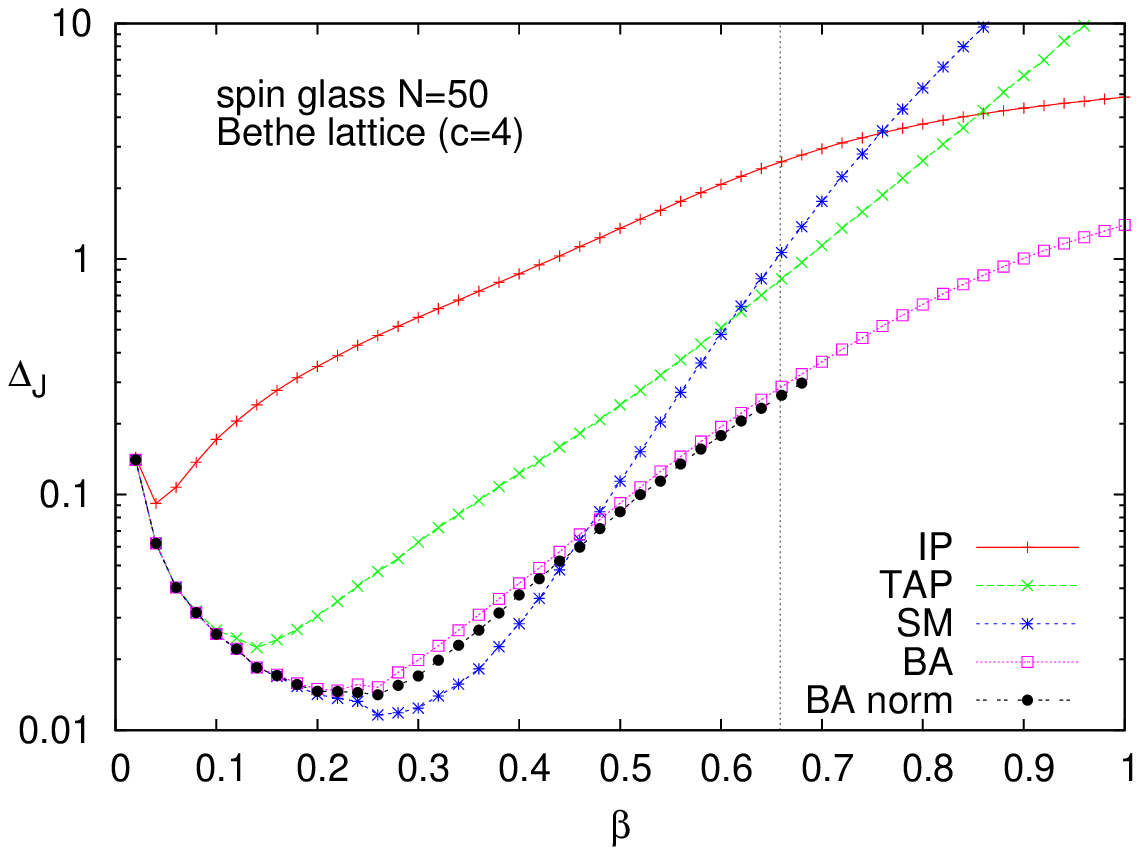}
\includegraphics[width=0.49\columnwidth]{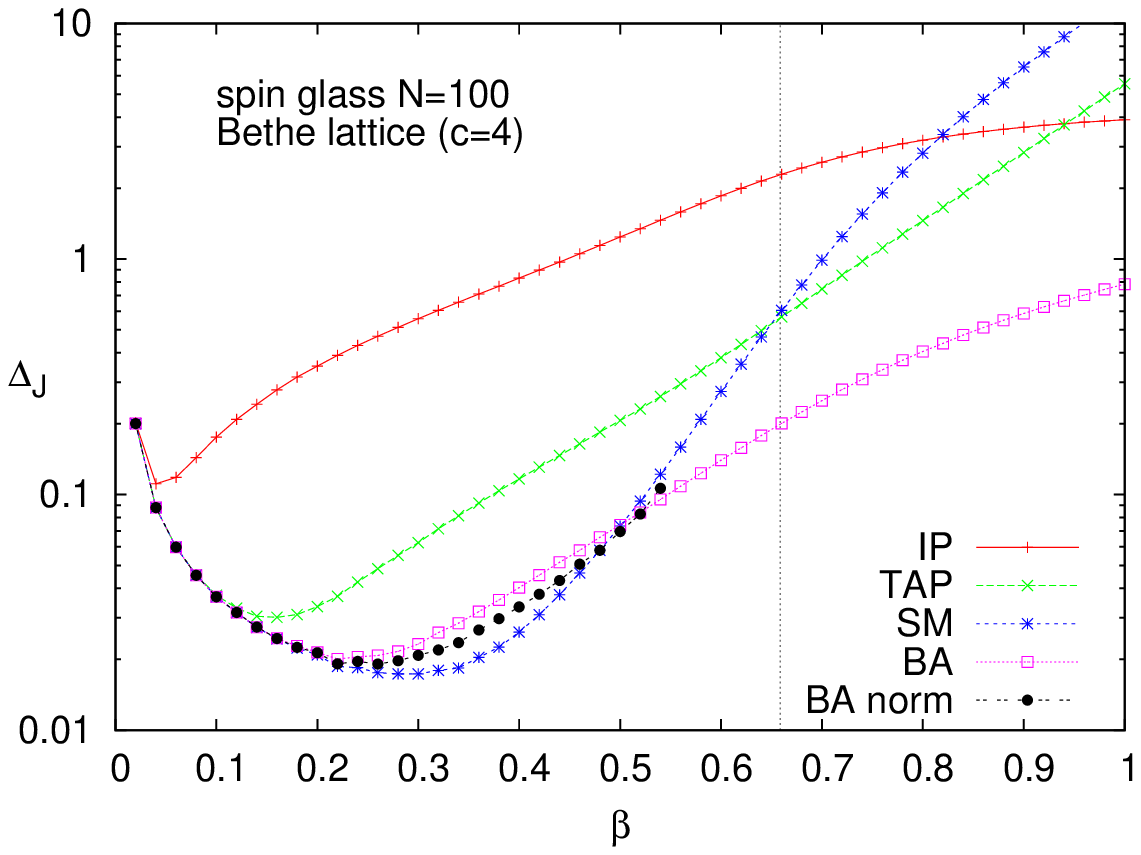}
\includegraphics[width=0.49\columnwidth]{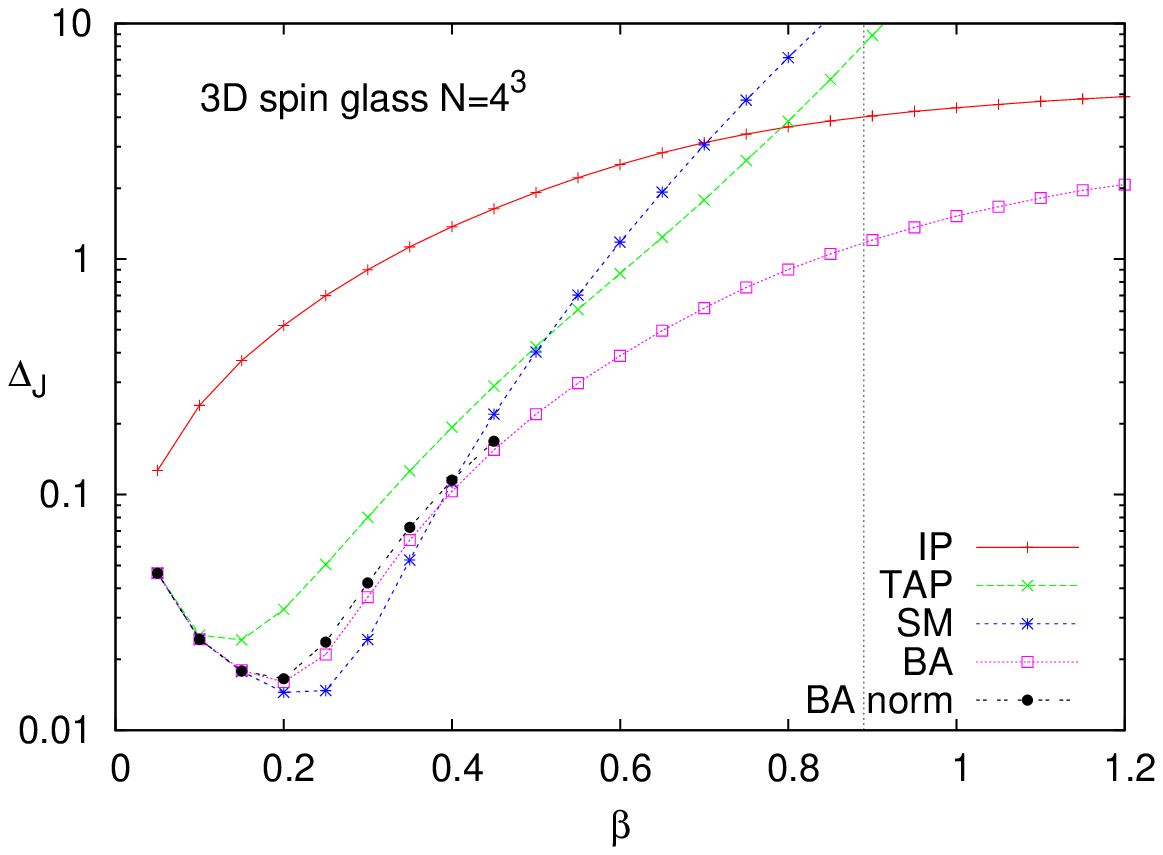}
\includegraphics[width=0.49\columnwidth]{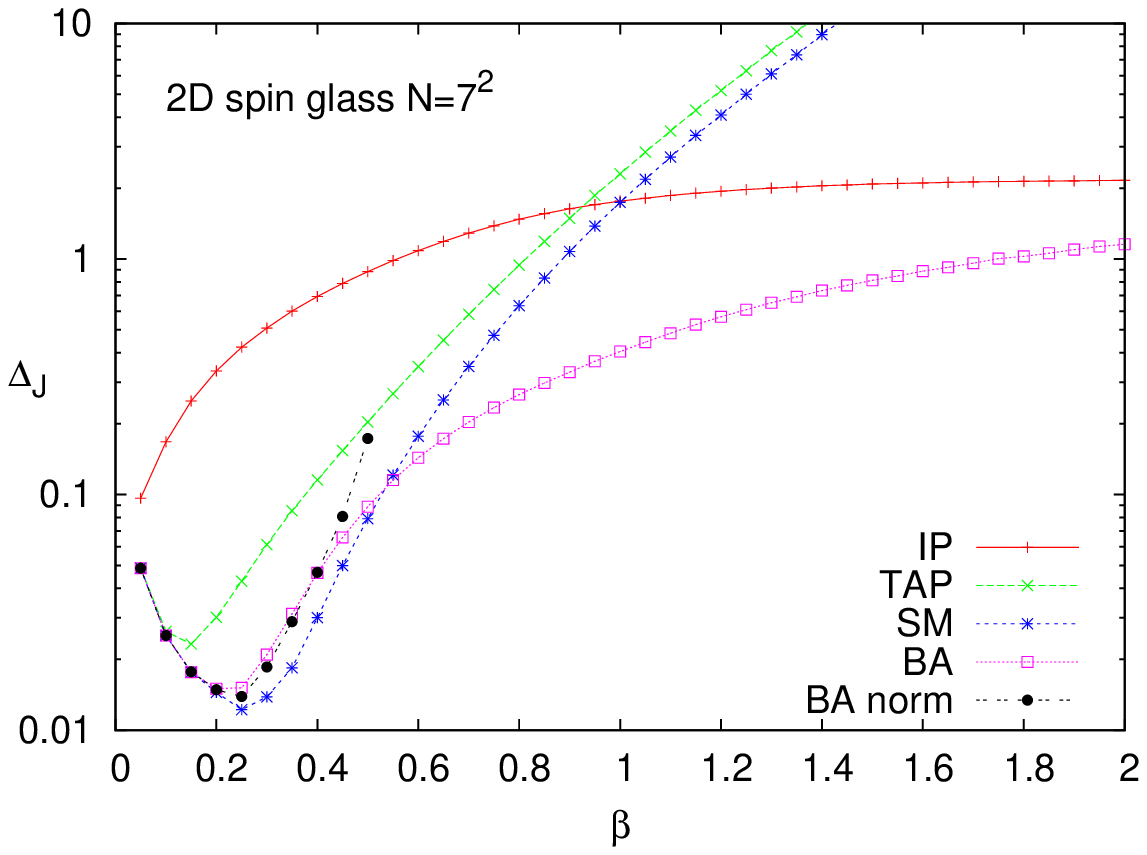}
\caption{Errors in the couplings inferred by several approximations for a typical sample of spin glass models ($J_{ij}=\pm\beta$) on different topologies: random regular graphs with fixed degree 4 (upper panels), 3D cubic and 2D square lattices (lower panels). The vertical dotted lines mark the loci of the spin glass phase transitions in the thermodynamic limit.}
\label{SGmodels}
\end{figure}

Also for spin glass models the conclusions reached above apply very well: the relative goodness of the various approximations is roughly unchanged. The only major difference is that the normalization trick does not provide any more a clear improvement: its performances strongly depend on the disorder sample. In Figure \ref{SGmodels} I am reporting the error on the inferred coupling for spin glass models on a random regular graph with fixed degree $c=4$, on a 3D cubic lattice and a 2D square lattice. Again, willing to suggest a general purpose inference method, the choice is clearly in favor of the BA.
Please notice that by running SuscProp it would be impossible to obtain the results shown in Figure~\ref{SGmodels}, because of the limited range of convergence of such an algorithm. Indeed for spin glass models on a random graph SuscProp stops converging around the critical temperature and for spin glass models on a regular lattice it stops converging even before, well into the high temperature phase: for example for the a spin glass model on a 2D square lattice it converges up to $\beta_\text{BP} \simeq 0.66$ \cite{GBP-GF} and on a 3D cubic lattice up to $\beta_\text{BP} \simeq 0.49$ \cite{Lage}. In this sense, the use of the new formula in Eq.(\ref{BA}) is really innovative.

\subsection{Spin glass models with an external field}

\begin{figure}
\includegraphics[width=0.49\columnwidth]{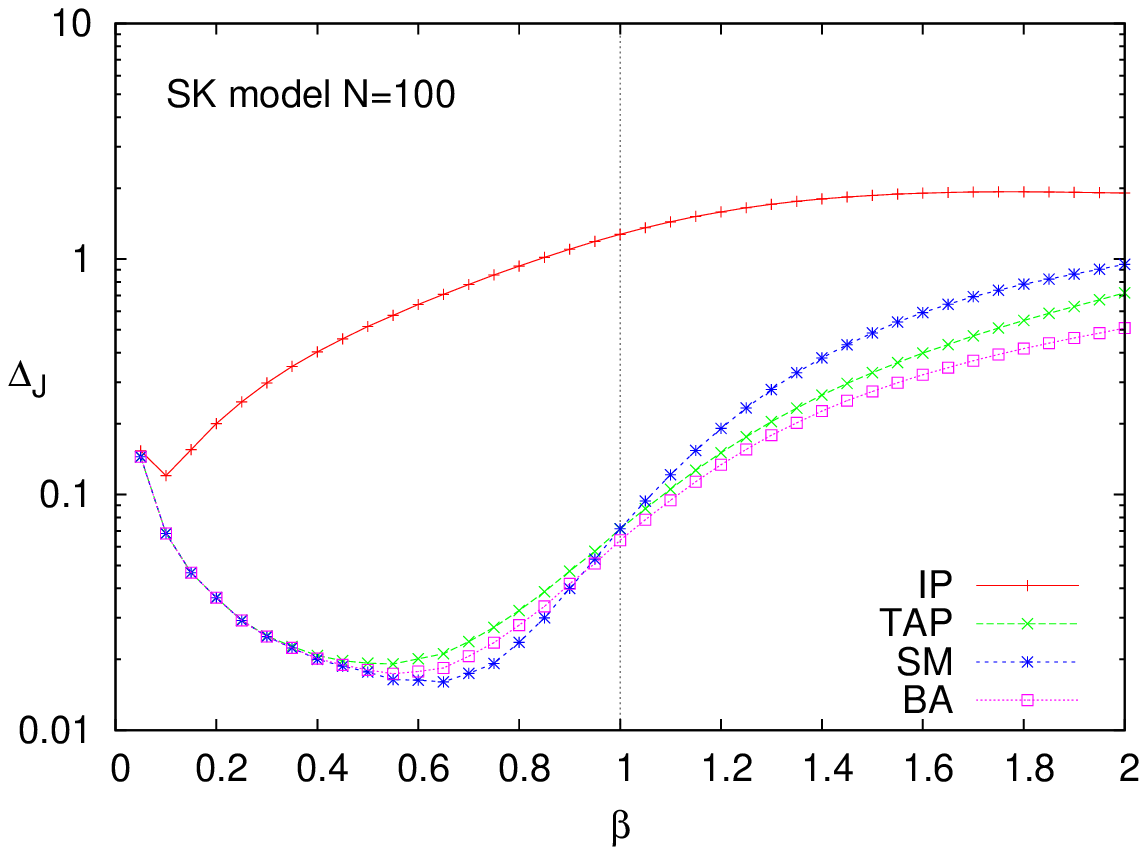}
\includegraphics[width=0.49\columnwidth]{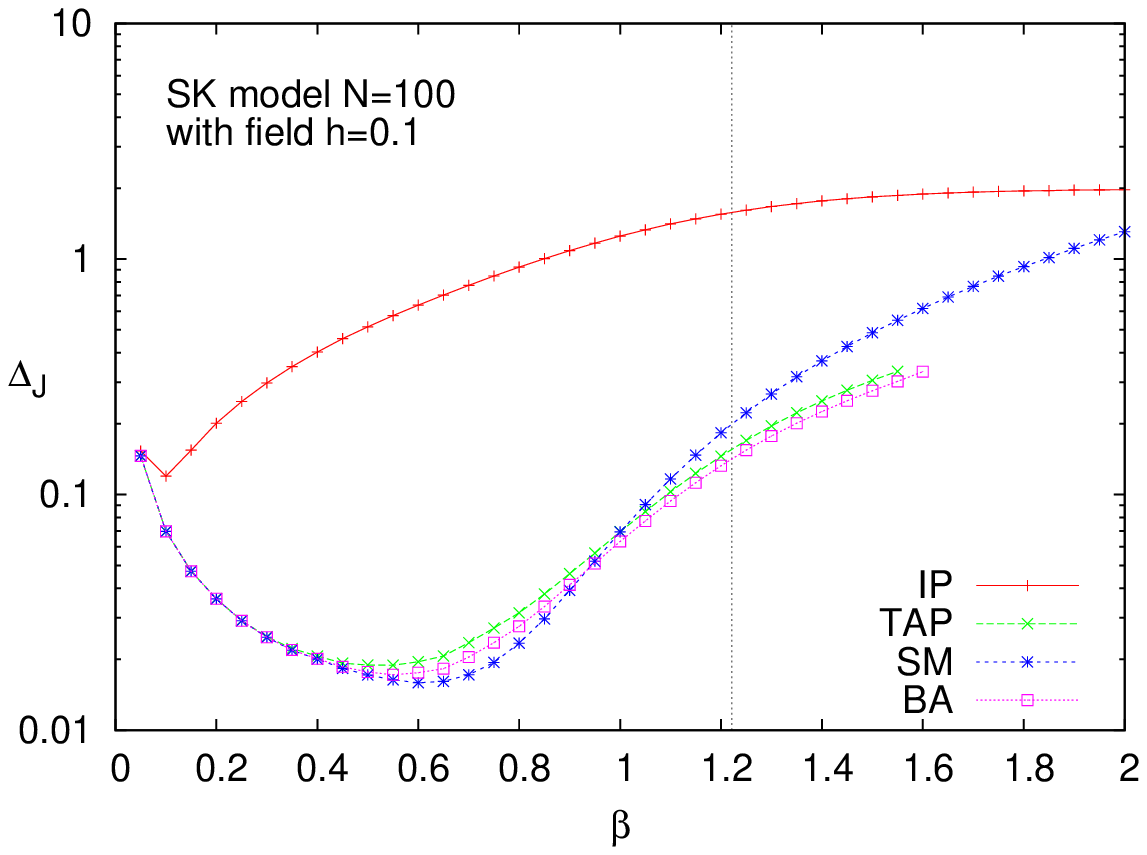}
\includegraphics[width=0.49\columnwidth]{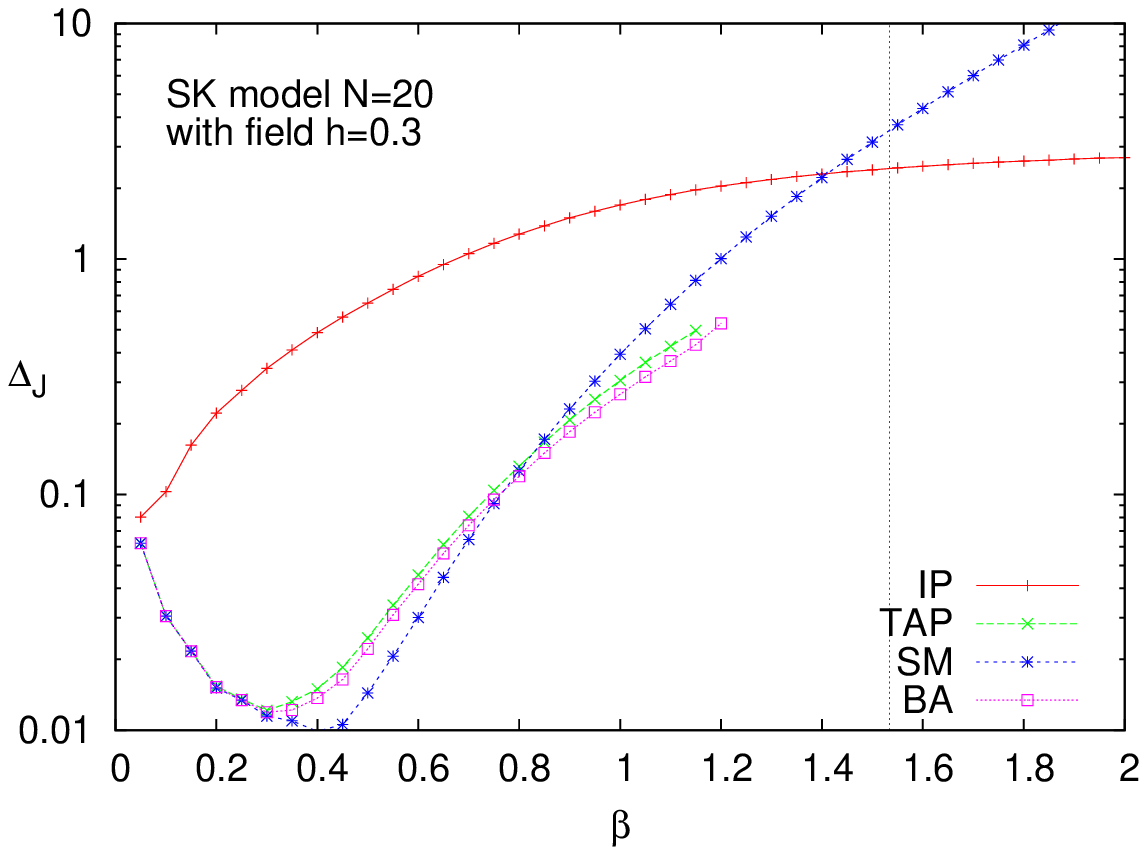}
\includegraphics[width=0.49\columnwidth]{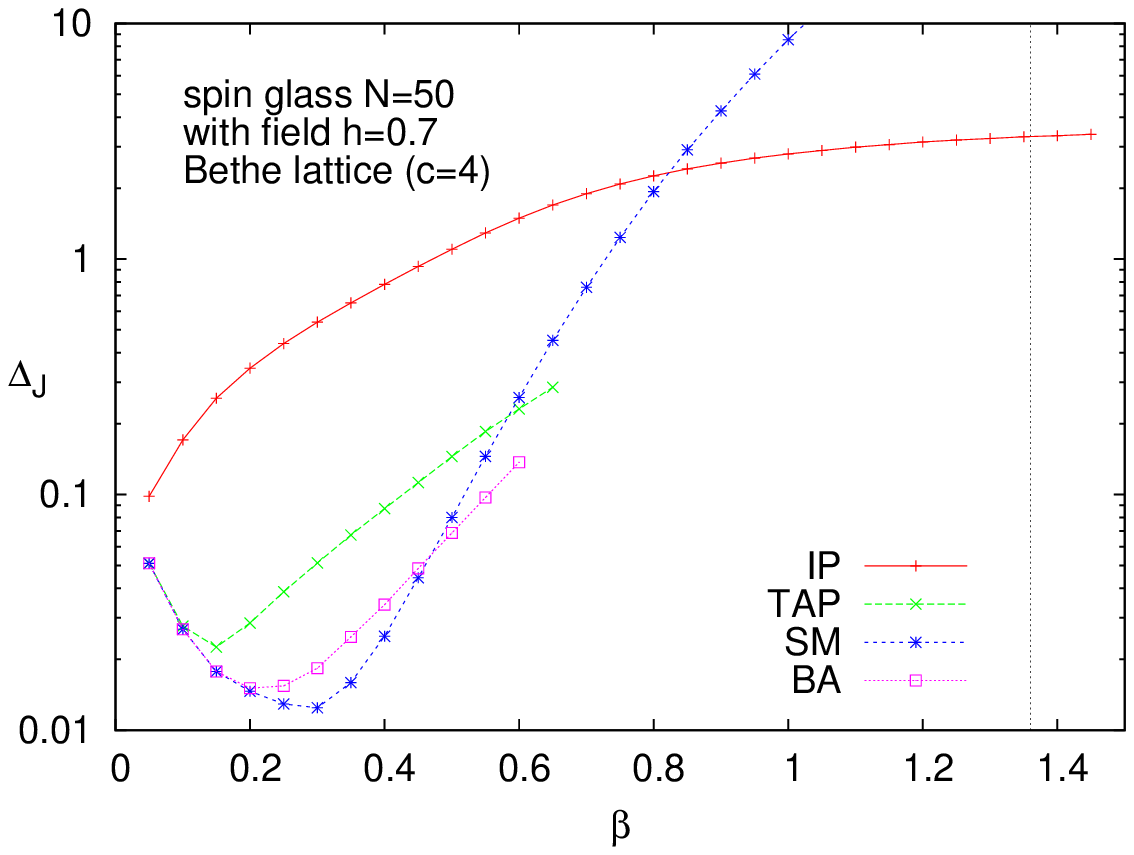}
\caption{Errors in the couplings inferred by several approximations. SK model $h=0$ (upper left, just for comparison), SK model $h=0.1$ (upper right), SK model $h=0.3$ (lower left) and spin glass on a random regular graph with fixed degree $c=4$ and $h=0.7$ (lower right). The vertical dotted lines mark the loci of the spin glass phase transitions in the thermodynamic limit. In presence of an external field TAP and BA cease to have a solution for high enough $\beta$ values.}
\label{SGfield}
\end{figure}

Let me finally come to the most surprising case: frustrated models in presence of an external field. As shown in Figure~\ref{SGfield}, once more the relative level of accuracy of the 4 approximations tested is the same, but the is a major difference with respect to case of zero field (which has been reported in the upper left panel of Figure~\ref{SGfield} for reader convenience). At high enough $\beta$ values, both TAP and BA cease to have a solution to Eqs.(\ref{eq:TAP}) and (\ref{eq:BA}) or equivalently the expressions under the square roots in Eqs.(\ref{TAP}) and (\ref{BA}) become negative. To my knowledge, this fact has been never noticed in the past, although the TAP approximation for inferring couplings is largely used.

In the Appendix I sketch the analytical solution to the simplest model showing this phenomenon, namely a system of 3 spins connected by antiferromagnetic couplings in presence of an external field: such an analytical solution should convince the reader that the phenomenon is not due to any numerical inaccuracy related to the complexity of the models studied here, but it can be mathematically proved in a very simple model.

The absence of solutions in TAP and BA is evident in Figure~\ref{SGfield} where the corresponding curves are interrupted at a $\beta$ value that becomes smaller for larger fields. Beyond the point where BA stops providing inferred couplings, one should resort to other inference methods. Unfortunately at that point both SM and IP already give quite large errors, that keep growing fast. So in practice none of the methods studied in the present work is valid for inferring couplings in a frustrated model in presence of an external field at low enough temperatures.

\section{Conclusions}

The purpose of the present work is to make a detailed comparison among several approximation for solving the inverse Ising problem, i.e., estimating coupling and fields form magnetizations and correlations.

After having explained how to derive the mean-field approximations based on the Plefka expansion (naive mean-field, TAP, Bethe approximations, etc...), I have ranked these approximations on the basis of how good they are in solving the direct problem (i.e., computing the correlations given the couplings). TAP and Bethe turned out to be in general the best approximations available.

Secondly I have derived the new analytical formula (\ref{BA}) for inferring couplings from magnetizations and correlations under the Bethe approximation. This formula allow to infer couplings without running the Susceptibility Propagation algorithm, thus avoiding all the serious problems related to the lack of convergence of such an algorithm.

After having summarized the formulas giving the inferred couplings for the 4 approximations tested (independent-pair, TAP, Bethe and the small correlation expansion of Ref.~\onlinecite{SessakMonasson}), I have introduced a trick that, normalizing the correlation matrix, improves the TAP and the Bethe approximations in case of models being unfrustrated or weakly frustrated.

Finally I have presented the results of the comparison among the 4 approximations for inferring couplings in diluted ferromagnetic models and spin glass models. I have used several different topologies: fully connected graphs, regular random graphs, 3D cubic lattices and 2D square lattices.

At the beginning of Section~\ref{sec:numeric} a list of general statements about the performances of these approximations in solving the inverse Ising problem is given. The bottom-line suggestion is to use the Bethe approximation, Eq.(\ref{BA}), eventually with the normalization trick if the model is weakly frustrated or unfrustrated.

In case of frustrated models with an external field (that is with non-zero magnetizations) I have found an important limitation for the TAP and Bethe approximations: at low enough temperatures these approximations stop having a solution and can not be used any more for solving the inverse Ising problem. This is a fundamental limitation, that take place also in very simple systems (see the Appendix) and that was not noticed before.

Moreover, when the Bethe approximation stops inferring couplings, the other methods already have a rather large error. So, in my opinion, it is still an open problem to find an approximation that, using only the correlation matrix, is able to solve the inverse Ising problem in a frustrated model with a field at low enough temperatures.

\begin{acknowledgments}

I acknowledge interesting conversations with A. Decelle, S. Franz, A. Pelizzola, I. Perez-Castillo and J. Raymond, and financial support by the Italian Research Minister through the FIRB Project No. RBFR086NN1 on ``Inference and Optimization in Complex Systems: From the Thermodynamics of Spin Glasses to Message Passing Algorithms''.

After the completion of the present manuscript I learnt of a manuscript by H.~C. Nguyen and J. Berg \cite{Johannes} showing a formula similar to Eq.(\ref{BA}).

\end{acknowledgments}

\appendix

\section{Limits of TAP and BA inference methods for a frustrated model in a field}

In this appendix I show explicitly that the formulas derived with TAP and BA for solving the inverse Ising problem do not always admit a solution for the case of frustrated models.
In order to simplify the computation I focus on the simplest model showing this problem, namely a system of 3 spins interacting with antiferromagnetic couplings of intensity $J<0$, in presence of an external field of intensity $h$, whose probability distribution is
\[
P(s_1,s_2,s_3) \propto \exp[J(s_1 s_2+s_2 s_3+s_3 s_1)+h(s_1+s_2+s_3)]\;.
\]
Thanks to the symmetries in the above measure, each spin has the same local magnetization $m(J,h)$ and each pair of spins has the same correlation $c(J,h)$.

When using the TAP approximation for the inverse problem one has to solve the following equation for each coupling $J_{ij}$,
\[
2 m_i m_j J_{ij}^2 + J_{ij} + (C^{-1})_{ij} = 0\;,
\]
and the above equation admit a solution only if its discriminant is non-negative:
\begin{equation}
\Delta^\text{TAP} \equiv 1 - 8 m_i m_j (C^{-1})_{ij} \ge 0\;.
\end{equation}

\begin{figure}
\includegraphics[width=0.6\columnwidth]{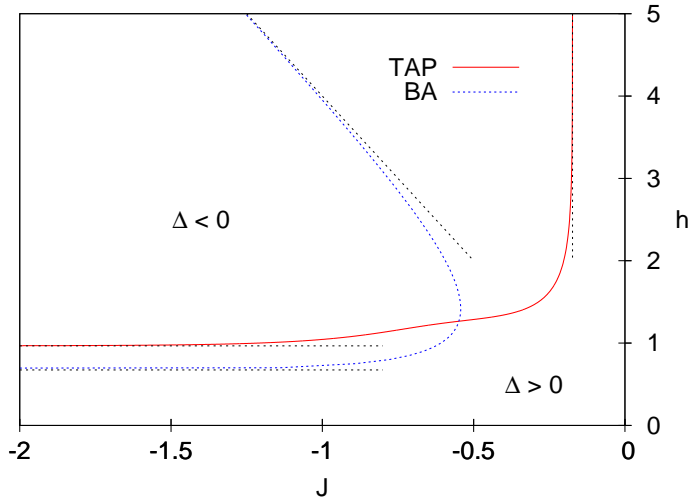}
\caption{Limit of validity of the TAP inference method for a system of 3 spins interacting with an antiferromagnetic coupling $J$ in an external field $h$. In the region where the discriminant $\Delta$ is negative the TAP inference method does not work.}
\label{fig:discr}
\end{figure}

In the present case, the discriminant is the same for each coupling and it is a function of the two parameters $J$ and $h$, that I report schematically in Figure~\ref{fig:discr}.
The full curve shown in Figure~\ref{fig:discr} corresponds to $\Delta^\text{TAP}(J,h)=0$ and has two asymptotes at $h^*=0.966759\ldots$ and $J^*=-\ln(2)/4$.
It is clear that for any non-zero field $h$ and any antiferromagnetic coupling $J$ the inference method based on the TAP approximation will fail at sufficiently small temperatures (i.e., large absolute values of $h$ and $J$).

The same phenomenon happens also for the inference method based on the BA. In this case the discriminant that may become negative is
\begin{equation}
\Delta^\text{BA} = \left(\sqrt{1+4(1-m_i^2)(1-m_j^2)(C^{-1})^2_{ij}}-2 m_i m_j(C^{-1})_{ij}\right)^2-4(C^{-1})^2_{ij}\;.
\end{equation}
In Figure~\ref{fig:discr} the dashed line corresponds to $\Delta^\text{BA}(J,h)=0$ and has two asymptotes at $h^*=0.673689$ and along the line $h=-4J$ (meaning that for this simple system of 3 spins the BA can work even a very low temperatures if the external field is large enough).

\end{document}